\documentclass[12pt,twoside]{article}
\usepackage[dvips]{graphicx}
\usepackage{float}
\usepackage{latexsym}
\usepackage{amsmath,mathrsfs,bm,amssymb,color}

\newtheorem{theorem}{Theorem}[section]
\newtheorem{proposition}[theorem]{Proposition}
\newtheorem{lemma}[theorem]{Lemma}
\newtheorem{corollary}[theorem]{Corollary}
\newtheorem{definition}[theorem]{Definition}
\newtheorem{example}[theorem]{Example}
\newtheorem{remark}[theorem]{Remark}

\newtheorem{assumption}[theorem]{Assumption}

\newcommand{\sg}{\sqrt{|{\rm det} g|}}
\newcommand{\AAA}{{\rm q}}
\newcommand{\rieman}{\eta}

\newcommand{\g}{\alpha}
\newcommand{\eb}{e^{-\theta(x)}}
\newcommand{\ms}[1]{\mathscr #1}
\newcommand{\bd}[1]{\begin{definition}\label{#1}}
\newcommand{\ed}{\end{definition}}
\newcommand{\la}{\lambda}
\newcommand{\bl}[1]{\begin{lemma}\label{#1}}
\newcommand{\el}{\end{lemma}}
\newcommand{\bc}[1]{\begin{corollary}\label{#1}}
\newcommand{\ec}{\end{corollary}}
\newcommand{\bt}[1]{\begin{theorem}\label{#1}}
\newcommand{\et}{\end{theorem}}
\newcommand{\bp}[1]{\begin{proposition}\label{#1}}
\newcommand{\ep}{\end{proposition}}
\newcommand{\br}[1]{\begin{remark}\label{#1}}
\newcommand{\er}{\end{remark}}
\renewcommand{\ggg}{f}
\newcommand{\vm}{v_{\rm m}}

\newcommand{\HH}{{\rm \iso }_0}
\newcommand{\iso}{\ms C}

\newcommand{\QQ}{L^2({\ms Q})}
\newcommand{\QQQ}{{L^2({\ms Q}_E)}}

\newcommand{\ov}[1]{\overline{#1}}
\newcommand{\dm}{d\mu_{\rm p}(x)}

\newcommand{\eq}[1]{\begin{equation}\label{#1}}
\newcommand{\en}{\end{equation}}
\newcommand{\eqn}{\begin{eqnarray*}}
\newcommand{\enn}{\end{eqnarray*}}
\newcommand{\eqnn}{\begin{eqnarray}}
\newcommand{\ennn}{\end{eqnarray}}
\newcommand{\proof}{{\noindent {\sc Proof}: \,}}
\newcommand{\qed}{\hfill $\Box$\par\medskip}
\newcommand{\BR}{{{\mathbb R}^3 }}
\newcommand{\bi}{\begin{description}}
\newcommand{\ei}{\end{description} }
\newcommand{\CC}{{{\mathbb C}}}
\newcommand{\RR}{{\mathbb R}}

\newcommand{\limn}{\lim_{n\rightarrow\infty}}
\newcommand{\limm}{\lim_{m\rightarrow\infty}}

\newcommand{\kak}[1]{(\ref{#1})}
\newcommand{\C}{\mathbb C}

\newcommand{\LR}{{L^2(\BR)}}
\newcommand{\hp}{H_{\rm p}}
\newcommand{\HP}{\ms H_{\rm p}}
\newcommand{\lp}{L_{\rm p}}
\newcommand{\lpb}{\bar L_{\rm p}}

\newcommand{\EE}{{\mathbb E}}
\newcommand{\ix}[1]
{\int \dm \EE^x\left[#1
\right]}
\newcommand{\ET}[1]{\EE_{\mu_T}\left[#1\right]}

\newcommand{\is}{\inf\sigma}
\newcommand{\f}{^{-1}}
\newcommand{\lk}{\left(}
\newcommand{\rk}{\right)}
\newcommand{\lkk}{\left\{}

\newcommand{\ab}[1]{\langle#1
\rangle}

\renewcommand{\d}{\displaystyle}

\newcommand{\add}{a^{\dagger}}

\newcommand{\MMM}[4]
{\left[ \!\!\!\begin{array}{cc}#1&#2\\
#3&#4\end{array}\!\!\!\right] }

\newcommand{\hf}{H_{\rm f}}

\newcommand{\hhh}{{\mathcal H}}
\newcommand{\gr}{\varphi_{\rm g}}
\newcommand{\grp}{\varphi_{\rm p}}
\newcommand{\grt}{\gr^T}

\newcommand{\half}{\frac{1}{2}}
\newcommand{\han}{{1/2}}

\newcommand{\WTT}{
\int_{-T}^T ds
\int_{-T}^T dt
W(X_s,X_t,|s-t|)
}

\newcommand{\WT}{
\int_{-T}^0 ds
\int_0^T dt W(X_s,X_t,|s-t|)
}

\newcommand{\wtt}{
\int_{-T}^T ds
\int_{-T}^T dt W}

\newcommand{\wt}{
\int_{-T}^0 ds
\int_0^T dt W}

\newcommand{\s}{\sigma}

\newcommand{\tvp}{\widetilde \chi}

\newcommand{\vp}{\chi}
\newcommand{\non}{\nonumber}
\newcommand{\wchi}{\check{\chi}}

\makeatletter
\@addtoreset{equation}{section}
\makeatother

\begin{document}

\title
{\sc
Infrared divergence of a scalar quantum field model
on a pseudo Riemannian manifold}
\author{
\small Christian G\'erard
\\[0.1cm]
{\small\it D\'epartment de Math\'ematique, Universit\'e de Paris XI}
 \\[-0.7ex]
{\small\it  91405 Orsay Cedex France}
\\[-0.7ex]
 {\small  {\tt christian.gerard@math.u-psud.fr}
 } \\[0.5cm]
\small Fumio Hiroshima\\[0.1cm]
{\small\it Department of Mathematics, University of Kyushu}
 \\[-0.7ex]
{\small\it  6-10-1, Hakozaki,
Fukuoka, 812-8581,  Japan}
\\[-0.7ex]
 {\small  {\tt hiroshima@math.kyushu-u.ac.jp}
 }
 \\[0.5cm]
\small Annalisa Panati
\\[0.1cm]
{\small\it D\'epartment de Math\'ematique, Universit\'e de Paris XI}
 \\[-0.7ex]
{\small\it  91405 Orsay Cedex France}
\\[-0.7ex]
{\small  {\tt annalisa.panati@math.u-psud.fr}}
\\[0.5cm]
\small Akito Suzuki
\\[0.1cm]
{\small\it Department of Mathematics, University of Kyushu}
\\[-0.7ex]
{\small\it  6-10-1, Hakozaki,
Fukuoka, 812-8581,  Japan}
\\[-0.7ex]
{\small  {\tt sakito@math.kyushu-u.ac.jp}}
}

\vspace{2cm}
\date{}
\pagestyle{myheadings}
\markboth{IR divergence}
{IR divergence}
 \setlength{\baselineskip}{18pt}
\maketitle
keywords: Nelson model, infrared divergence, pseudo Riemannian manifold, \\ functional integrals, ground states
\vspace{1cm}
\begin{abstract}
\noindent
A scalar quantum field model
defined on  a pseudo Riemannian manifold
is considered.
The model is unitarily transformed to the one with a variable mass.
By means of a Feynman-Kac-type formula,
it is shown that
 when the  variable mass is short range, the Hamiltonian
has no ground state.
Moreover
the infrared divergence
 of the expectation values of the number of bosons in the ground state is discussed.
\end{abstract}

\section{Introduction}
\subsection{Preliminaries}
Analysis of the infrared behavior in massless quantum field
theory is an important issue.
The infrared divergence
is seen to arise as follows:
the emission probability
of {\it massless} boson
becomes infinite
with increasing wavelength.
For some scalar quantum field model,
which is the so-called Nelson model
\cite{nel64},
a sharp result concerning
the relationship between
the infrared behavior and the existence (or the absence)
of ground states
is known.
The Nelson model describes
a scalar field coupled to a quantum mechanical particle with external potential $V$
in such a way that the interaction is linear.
Namely the Nelson model with mass $m_0\geq 0$
is formally given by
 \eq{n1}
H_{\rm N}=\half p^2+V(q)+
\half
\int \lk
\pi(x)^2+(\nabla\phi(x))^2+
m_0^2\phi(x)^2
\rk
dx+\int\phi(x)\vp(x-q) dx,
\en
where $\chi$ denotes a cutoff function,
$p$ and $q$ are the position operator and momentum operator of the particle, respectively,  with bare mass $1$,
which satisfy
$[p,q]=-i$,
and
$\pi(x)$ is the momentum field canonically conjugate to the scalar field $\phi(x)$,
which satisfy
$[\phi(x),\pi(y)]=i\delta(x-y)$.
 The dispersion relation for the Nelson model is given by
\eq{nd}
\widehat\omega_{\rm N}=\sqrt{-\Delta +m_0^2}
\en
in the position representation
and
 the
equation of motion is
\begin{eqnarray}
\label
{vn1}
&&
(\square
  +m_0^2 )\phi(x,t) = -\chi(x-{q_t}),\\
&&
\label{vn2}
\partial_t^2 q_t=-\nabla_q V(q_t)-
\nabla_q \phi(\chi(x-{q_t})),
\end{eqnarray}
where $\square =\partial_t^2-\Delta_x$.
It is established that
$H_{\rm N}$
with positive mass $m_0>0$
has a ground state
but
no ground
state for
$m_0=0$,
and the expectation value of the number of bosons in the ground state diverges as $m_0\rightarrow 0$.

While
the Nelson  model
defined on a {\it static} Riemannian manifold
is unitarily transformed to
a model with
 a variable mass
\eq{k8}
\vm (x) = m(x)^2 \geq 0
\en
and the dispersion relation \kak{nd}
is changed to
\eq{101}
\widehat\omega=\sqrt{-\Delta +
\vm. }
\en
By comparing \kak{nd} and \kak{101},
the variable mass is
seen to intermediate between massive
cases and massless cases,
and furthermore the infrared behavior,
as mentioned below,
depends on the decay property
of $\vm (x)$ as $|x|\rightarrow \infty$.

We consider in this paper a version of the Nelson model with variable masses.
 The  Hamiltonian is formally given by \eq{k1}
H_{\rm formal}=\half p^2+V(q)+
\half
\int \lk
\pi(x)^2+(\nabla\phi(x))^2+\vm (x)\phi(x)^2
\rk
dx+\g\phi(\rho_q),
\en
where $p$ and $q$,
 and
 $\phi(x)$ and $\pi(y)$
 satisfy the same canonical commutation relations as that of the Nelson model.
 The field operator $\phi(\rho_q)=\int \phi(x)\rho_q(x) dx$ is, however,
a scalar field smeared
by some  function $\rho_q$ defined through $\vm $ and a given cutoff function $\chi$,
and
 $\g$ a real coupling constant.
Thus the
equation of motion is
given by
\begin{eqnarray}
\label
{v1}
&&
(\square
  +\vm (x) )\phi(x,t) = -\g\rho_{q_t}(x),\\
&&
\label{v2}
\partial_t^2 q_t=-\nabla_q V(q_t)-\g\nabla_q \phi(\rho_{q_t}).
\end{eqnarray}
Here $\square  +\vm(x)$ appears in \kak{v1}
instead of $\square  +m_0^2$.
This is a unitary transformed version of a Klein-Gordon equation  defined on a pseudo Riemannian manifold. See Section 2.5.

We are interested in investigating the infrared behavior
of the Nelson model.
\begin{figure}[ht]
    \centering
    \includegraphics{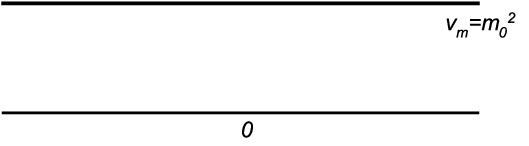}
    \caption{Positive constant mass}
    \label{ho1}
\end{figure}
In the case of constant mass $\vm(x)=m_0^2$ in \kak{101},
it is established that
if $m_0 > 0$,
the Nelson model has the unique ground state up to
multiple constants
(Fig.\ref{ho1}),
but if $m_0=0$
no ground state exists unless
    the  infrared regularization
    is imposed.
    See e.g., \cite{bfs2,bhlms, che01,ger00,hh06,hi06,lms,spo98}
for detail.
Here
the infrared regular condition is defined by
\eq{k6}
\int_{\BR}
 \frac{\vp(k)^2}{|k|^3}dk<\infty.
\en
Conversely
\eq{k7}
\int_{\BR}
 \frac{\vp(k)^2}{|k|^3}dk =\infty
\en
is called the infrared singular condition. The singularity in \kak{k7} comes from
a neighborhood of $k=0$ if $\chi$ has a compact support, since the dimension is three.

\begin{figure}[ht]
    \centering
    \includegraphics{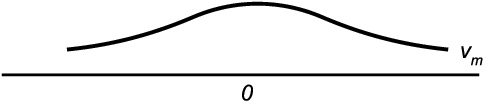}
    \caption{Long range variable mass}
    \label{ho2}
\end{figure}

\begin{figure}[ht]
    \centering
    \includegraphics{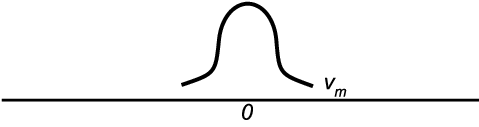}
    \caption{Short range variable mass}
    \label{ho3}
\end{figure}

  Our paper is motivated by extending constant mass cases to variable ones. Namely, going beyond the case of constant masses,
we consider the infrared behavior of the Nelson model with variable masses.
From the argument mentioned above it
is expected that the Nelson model
may have ground states if the variable mass
decays sufficiently slowly in a neighborhood of origin (Fig. \ref{ho2}),

\noindent
but no ground state exists if it decays sufficiently fast (Fig. \ref{ho3}).
Taking into account of this intuitive  argument, as the first step,
we consider two cases:
(1)
$\vm $ is long range
and
(2)
$\vm $ is short range.
In this paper we focus on (2) and
prove that
for a short range potential $v\geq 0$ such that
$
\vm (x)=\mathcal O(|x|^{-\beta})$
with
$
\beta>3$,
$H$ has no ground state in the Hilbert space unless the infrared regularization is imposed.

\subsection{Strategy}
It is proven  that
the functional integration is useful device to show
the existence and non-existence of
the ground state of the Nelson model
with constant masses.
It can be extended to the case of
variable masses in this paper.
The main tool used in this
 paper is
 functional integral representations of the semigroup $e^{-tH}$ and
 an extension of the strategy developed in \cite{bhlms, lms} where the Nelson model with constant mass is discussed.

The Nelson model  $H$ can be defined as a self-adjoint operator
 on some probability space.
It is easily shown that
\eq{k2}
\grt=\|e^{-TH}1\|\f e^{-TH}1,\quad T>0,
\en
 is a sequence approaching to a ground state of $H$ {\it if a ground state exists}.
Conversely
\eq{k5}
\lim_{T\rightarrow \infty}
 (1,\grt)^2=a>0,
\en
implies
the existence of
the  ground state of $H$,
but the absence of ground state
follows
from
\eq{102}
\lim_{T\rightarrow \infty}
 (1,\grt)^2=0.
\en
By making use of a modification of \cite{lms}
we show that \kak{102} holds under the infrared singularity condition \kak{k7}.

Throughout this paper we use the notation
$\mathbb E_\mu[\cdots]$ for $\int \cdots d\mu$ and $\mathbb E_\nu^x[\cdots]$ for $\int\cdots d\nu^x$,
where $\nu^x$ denotes a probability measure starting at $x$ on
a path space.
By using
the functional integration,
we have the bound
\eq{k3}
(1,\grt)^2\leq \ET{e^{-\g^2\WT}}
\en
with some probability measure $\mu_T$ on the product configuration
space $\BR\times C(\RR;\BR)$
and the so-called double potential
$W=W(X_s, X_t, |s-t|)$ given by
\eq{do}
W(X,Y,|t|)=\int
\frac{\vp(k)^2}{2|k|}
\ov{\Psi(k,X)}{\Psi(k,Y)}e^{-|t||k|}dk.
\en
Here $\Psi(k,x)$ denotes the generalized eigenvector
of $-\Delta +\vm $.
By controlling the behavior of measures $\mu_T$ and $\WT{}{}$
as $T\rightarrow \infty$,
we can show \kak{102} under the infrared singular condition.

Next we consider the expectation values of the number of bosons in the ground state $\gr$.
Assume the infrared regular condition \kak{k6} and the existence of ground state.
Let $N$ be the number operator.
  We can show that
$(\grt, e^{-\beta N} \grt)$ can be analytically continued from $\beta\in [0,\infty)$ to the whole complex plane
$\beta\in\CC$.
Then the moment $(\grt, N^n\grt)$ is given by
$$(\grt, N^n\grt)=
\left.
(-1)^n \frac{d^n}{d\beta^n}
(\grt, e^{-\beta N}\grt)\right \lceil_{\beta=0}.$$
As an application
we can show that
the expectation value of the number of bosons in the ground state,
$(\gr, N\gr)$, diverges as
$\int_{\BR} \frac{\vp(k)^2}{|k|^3}dk$
 tends to infinity.

This paper is organized as follows:
Section 2 is devoted to giving  the definition of the Nelson model with a variable mass.
In Section 3 we discuss functional integration in Euclidean quantum field theory.
In Section 4 we prove the absence of ground state. Finally in Section 5 we show the divergence of $(\gr, N\gr)$ in infrared singularity.

\section{The Nelson model on a pseudo Riemannian manifold}
\subsection{Particle}
We introduce the Schr\"odinger operator $\hp$ by
\eq{3}
\hp=\half p^2+V,
\en
where $p_\mu=-i\nabla_\mu$, $p^2=p\cdot p$,  and $V$
is an external  potential.
We say that
$V$ is Kato-class  if and only if
$$\lim_{r\downarrow0}\sup_{x\in\BR}\int_{|x-y|<r}
\frac{|V(y)|}{|x-y|}dy=0
$$
and $V$ is  local Kato-class  if and only if
$1_K V$ is Kato-class  for arbitrary compact set $K\subset \BR$.
If $V=V_+-V_-$ satisfies that $V_+$ is local Kato-class and $V_-$ Kato-class, we say that $V$ is Kato-decomposable.
When $V$ is Kato-class,
$V\in L_{\rm loc}^1(\BR)$
and $V$ is infinitesimally small with respect to $p^2$ in the sense of form,
furthermore when $V=L^p(\BR)+L^\infty(\BR)$ with $p> 3/2$, $V$ is Kato-class.
In particular an arbitrary polynomial is local Kato-class.

We introduce assumptions on external potential $V$:
 \begin{assumption}\label{80}
{\bf (Assumptions on $V$)}
We assume (1)-(3) below:
\bi
\item[(1)]
  $V=V_+-V_-$ is Kato-decomposable
  with $V_-\in L_{\rm loc}^p(\BR)$ for some
   $p> 3/2$.
   \item[(2)]
$V$ is bounded from below and   $V(x)>C|x|^{2\AAA}$ with some $\AAA >0$ for $x\in \BR\setminus M$ with some compact set $M$.
\item[(3)]
The ground state  of $\hp$ is unique and strictly positive.
 \ei
  \end{assumption}
$\hp$ is defined as a quadratic form sum.
Since $V$ is Kato-decomposable,
 $\hp$   is closed on $Q(p^2)\cap Q(V_+)$ and bounded from below, where $Q(T)$ denotes the form domain of $T$.
  See \cite[Theorem A.2.7]{sim}.
Moreover it follows that
$
\sup_{x\in\BR} \mathbb E_{P_W}\left[
e^{-\int_0^tV(B_s+x)ds}
\right]
<\infty
$
for arbitrary $t\geq0$, where $(B_t)_{t\geq0}$ denotes the $3$-dimensional Brownian motion starting at zero on a probability space
$(W, \ms B_W, P_W)$.
By (2) of Assumption \ref{80},  $V\to \infty$ as $|x|\to \infty$. Then $\hp$ has a compact resolvent.
This can be proven  by showing
that $\{\psi\in Q(\hp)|\|\psi\|\leq 1,
(\psi,\hp\psi)\leq1\}$ is compact in $\LR$.
See e.g., \cite[Theorem XIII.67]{rs4}.
In particular the spectrum of $\hp$ is purely discrete and the ground state $\grp$ of $\hp$ exists.
By assumptions,
$V_+\in L_{\rm loc}^1(\BR)$ and
$V_-\in L^p(\BR)$ with $p> 3/2$, and $V(x)>C|x|^\AAA $ for sufficiently large $|x|$,
it is known that $\grp(x)$ exponentially decays.
We used this in Section 4.

Now let  us define a unitary transformation.
By (3) of Assumption \ref{80}
we can define the ground state transformation $$U_{\rm p}:\LR\rightarrow \HP=L^2(\BR,\grp^2 dx)$$ by
\eq{4}U_{\rm p}f=\frac{1}{\grp}f.
\en
Set
\eq{6}
\lp=U_{\rm p}\hp U_{\rm p}\f
\en
and the probability measure $\mu_{\rm p}$ on $\BR$
is defined by
\eq{meas}
\dm =\grp^2(x)dx.
\en
Thus the operator $\lp$ acts on the {\it probability} space $L^2(\BR;d\mu_{\rm p})$.
Formally
$\lp$ is given by
\eq{lp}
\lp f= -\half \Delta f+\frac{\nabla \grp }{\grp}\nabla f
\en
on $L^2(\BR;d\mu_{\rm p})$, it is
  of course not clear whether $\grp\in C^1(\BR)$ or not. However
by the  Kolmogorov consistency theorem
 we can construct a continuous Markov process $X=(X_t)_{t\in\RR}$ associated with the semigroup
$e^{-t\lp}$.
 This process $X$ is a formal solution of the stochastic differential equation:
$$dX_t=dB_t+
  \frac{\nabla \grp }{\grp}(X_t)dt.$$
We will discuss the Markov process $X$
in Section 3.

\subsection{Boson Fock space}
The Boson Fock space over
 the one particle space $\LR$ is defined
 by
$$\ms F=\bigoplus_{n=0}^\infty
L_{\rm sym}^2(\RR^{3n}),$$
 where $L_{\rm sym}^2(\RR^{3n})$
 is the set of $L^2$ functions $f(k_1,...,k_n)$, $k_j\in\BR$, $j=1,...,n$,
 on $\RR^{3n}$ such that it is symmetric with respect to $k_1,...,k_n$ with
  $L^2_{\rm sym}(\RR^0)=\C$.
 The Fock vacuum
 $1\oplus 0\oplus 0\oplus\cdots$
  in $\ms F$
 is denoted by $\Omega_{\ms F}$.
The annihilation operators $a(f)$
smeared by $f\in\LR$
and the creation operators $\add(g)$ by $g\in\LR$
are defined in $\ms F$ and satisfy canonical
commutation relations:
\begin{eqnarray}
&&
[a(f),\add(g)]=(\bar f, g)_{\LR},\\
&&
[a(f),a(g)]=0=[\add(f),\add(g)].
\end{eqnarray}
Here $(f,g)_{\ms K}$ denotes the scalar product
on a Hilbert space $\ms K$. We omit $\ms K$ unless confusion arises.
Note that
$$(a(f))^\ast=\add(\bar f)$$
and that
$\add(f)$ and $a(f)$ are linear in $f$.
We formally write $a(f)=\int a(k) f(k)dk$ and $\add(f)=\int \add(k) f(k) dk$.
For a contraction
operator $T:\LR\rightarrow \LR$,
define the contraction operator $\Gamma(T):
\ms F\rightarrow \ms F$ by
$\Gamma(T)\Omega_\ms F= \Omega_\ms F$
and
$$\Gamma(T)\add(f_1)\cdots\add(f_n)
\Omega_\ms F=
\add(T f_1)\cdots \add(Tf_n)\Omega_\ms F.$$
Note that $\Gamma(TS)=\Gamma(T)\Gamma(S)$ and $\Gamma(I)=I$.
Then
for a self-adjoint operator
 $h$ in $\LR$ there exists a unique
 self-adjoint operator $d\Gamma(h)$
in $\ms F$   such that
$$e^{itd\Gamma(h)}=\Gamma(e^{ith}),\quad
t\in\RR.$$

\subsection{The Nelson model with variable mass}
Let us assume that $-\Delta+\vm $ is a self adjoint operator in $L^2(\mathbb{R}^3)$.
Suppose that $-\Delta +\vm $ has generalized eigenfunctions $\Psi(k,x)$:
 \eq{92}
 (-\Delta+\vm (x))\Psi(k,x)=|k|^2 \Psi(k,x),\quad k\in\BR.
 \en
We introduce the following assumptions.
\begin{assumption}
\label{ge}
{\bf (Assumptions on $\Psi(k,x)$)}
The generalized eigenvectors
satisfy that
\bi
\item[(1)]
  $\sup_{k,x}|\Psi(k,x)|<\infty$,
   \item[(2)]
   $\Psi(k,x)$ is continuous in $x$ for almost every  $k$,
\item[(3)]
the generalized Fourier transformation:
\eq{94}
(\mathcal F f)(k)=(2\pi)^{-3/2}
{\rm l.i.m.}\int f(x) \ov{\Psi(k,x)}dx
\en
is unitary on $\LR$.
   \ei
\end{assumption}
By (3) above  the inverse of $\mathcal F$,
 ${\mathcal F}\f$,  is given by
\eq{94-1}
(\mathcal F^{-1} g)(x)=(2\pi)^{-3/2}
{\rm l.i.m.}\int g(k) {\Psi(k,x)}dk.
\en
Recall that $\widehat\omega=\sqrt{-\Delta+\vm }$.
Then we have
\eq{95}
\mathcal F\widehat   \omega \mathcal F\f =  \omega,
\en
where
 $\omega$ is the multiplication operator given  by
\eq{40}
\omega(k)=|k|,\quad k\in\BR.
\en
Let $\chi$ be a cutoff function.
We define
the field operator with the variable mass $\vm $ and
the  cutoff function $\chi$ by
\eq{94*}
\widehat\Phi(x)=\frac{1}{\sqrt 2}
\lk
\add
\lk
{\widehat \omega^{-\han}\rho_x}
\rk
+a
\lk
\ov{
\widehat \omega^{-\han}\rho_x}
\rk\frac{}{}\!\!\rk,
\en
where
\eq{rho}
 \rho_x(\cdot)
		= (2\pi)^{-3/2}
\int
\Psi(k,\cdot)
\ov{\Psi(k,x)}
\vp(k)dk.
		\en
A physically reasonable choice
of $\chi$ is
\eq{k100}
 \vp(k) = \frac{\chi_\Lambda(|k|)}{\sqrt{(2\pi)^3}}, \quad \Lambda >0,
 \en
 where $\chi_\Lambda$ is an ultraviolet cutoff defined
 by
$
 \chi_\Lambda(s) = \begin{cases} 0, & s \geq \Lambda \\
							1, & s < \Lambda \end{cases} $.
If we take $\kak{k100}$ as $\chi$,
then $\rho_x \rightarrow \delta(\cdot -x)$ in $\mathscr{S}'$ as $\Lambda \to \infty$.

Let us define the free Hamiltonian
$\widehat {\hf}$ by
\eq{k11}
\widehat{\hf}=d\Gamma(\widehat \omega).
\en
The total state space is defined
by the tensor product of $\hhh_{\rm p}$ and $\ms F$:
\eq{k18}
\mathcal H=\hhh_{\rm p} \otimes\ms F.
\en
\begin{definition}
{\bf (The Nelson model with variable mass)}
{\rm
The Nelson Hamiltonian
with the variable mass $\vm $ is defined by
\eq{k13}
\widehat H=\lp\otimes 1+
1\otimes \widehat {\hf}+\g\widehat{\Phi}
\en
on the Hilbert space $\hhh$, where
$\widehat \Phi=\int^\oplus_\BR \widehat\Phi(x) dx$ under the identification $\hhh=\int^\oplus_\BR \ms F ds$.
}
\end{definition}
Now we derive the equation of motion associated with $\widehat H$.
Let
\eq{k19}
 \varphi(f) = \frac{1}{\sqrt{2}}
\lk
\add
\lk
{\widehat \omega^{-\han}f}
\rk
+a
\lk
\ov{\widehat \omega^{-\han}f}
\rk\frac{}{}\!\!\rk
\en
be the field operator
smeared by $f$.
Then
$\widehat \Phi(x)=
\varphi(\rho_x)$.
The time evolution
of $\varphi(f)$ is given by
\eq{k14}
\varphi(f,t) =
e^{it \widehat H}
\varphi(f)e^{-it \widehat H}
\en
and that of $x$ by
\eq{k14-1}
q_t=
e^{it \widehat H}
x e^{-it \widehat H}.
\en
Since
$$
[d\Gamma{(\widehat \omega)},a(f)]=-a(\widehat \omega f),
\quad
[d\Gamma{(\widehat \omega)},\add(f)]=\add (\widehat \omega f),
$$
$\varphi(f,t)$ and $q_t$ satisfy  that
\begin{eqnarray}
&&
\partial_t^2
\varphi(f,t)
+
\varphi((-\Delta +\vm ) f,t)=-\g(\rho_{q_t},f),\\
&&
\partial_t^2 q_t
=-\nabla V(q_t)-\g\varphi(\nabla \rho_{q_t})
\end{eqnarray}
on $\mathcal H$. Compare with \kak{v1} and \kak{v2}.

\subsection{Unitary transformation}
In this subsection we unitarily transform
the Nelson Hamiltonian
to some self-adjoint operator $H$.
Let
$\hf $ be defined by
\eq{34-1}
\hf =d\Gamma(\omega)
\en
and
$\Phi(x)$ by
\eq{1}
\Phi(x)=\frac{1}{\sqrt 2}
\int
\lk
\frac{\vp(k)}{\sqrt{\omega(k)}}
\ov{\Psi (k,x)} \add(k)+
\frac{{\vp(k)}}{\sqrt{\omega(k)}}
{\Psi (k,x)} a(k)\rk
dk.
\en
Define
$H$ by
\eq{97}H=\lp\otimes 1+1\otimes\hf+\g\Phi,
\en
where $\Phi=\int^\oplus_\BR\Phi(x) dx$.
We introduce some assumption on cutoff function $\chi$.
\begin{assumption}
\label{sa}
{\bf (Assumptions on $\vp$)}
Assume that $\vp$ is real,
 $\wchi\geq0$ $(\not=0)$,
$\vp/\sqrt\omega\in\LR$ and
$\vp/\omega\in\LR$,
where $\wchi$ denotes the inverse Fourier transform of $\vp$.
\end{assumption}
\begin{remark}
Since the space dimension under consideration is three,
 from $\wchi\geq0$ in Assumption \ref{sa} it follows that $\vp(0)>0$ and then
 it follows that
 \eq{IR}\int \frac{\chi(k)^2}{\omega(k)^3}dk=\infty.
\en
\end{remark}
The next proposition is standard.
\bp{sa1}
Suppose Assumption \ref{sa} and (1) of Assumption \ref{ge}.
Then
the Nelson Hamiltonian
$H$ (resp. $\widehat H$) is
self-adjoint on $D(\lp)\cap D(\hf)$
(resp. $D(\lp)\cap D(\widehat{\hf})$~)
and bounded from below. Moreover
$H$ (resp. $\widehat H$) is essentially self-adjoint on any core of
$\lp\otimes 1+1\otimes \hf$ (resp.
$\lp\otimes 1+1\otimes \widehat{\hf}$).
\ep
\proof
Since $\Phi$ (resp. $\widehat\Phi$) is infinitesimally
small with respect to $\lp\otimes 1+1\otimes \hf$ (rep. $\lp\otimes 1+1\otimes \widehat{\hf}$),
the proposition follows from the
Kato-Rellich theorem.
\qed
Let $\mathcal F_b=\Gamma(\mathcal F)$ which
is a unitary operator on $\ms F$.
\bp{k15}
Suppose Assumption \ref{sa} and (1) of Assumption \ref{ge}.
Then
\eq{u}
 H  =(1\otimes \mathcal F_b)\widehat H (1\otimes \mathcal F_b\f).
 \en
 \ep
\proof
Since
$$\mathcal F \hat
\omega^{-1/2}
\rho_x(\cdot) \mathcal
=\omega^{-\han}(\cdot)\chi(\cdot) \ov{\Psi(\cdot,x)}$$ and
$\mathcal F_b \add(\widehat \omega^{-\han}\rho_x)\mathcal F_b^{-1}=\add(\mathcal F \widehat \omega^{-\han}\rho_x)$
and
$\mathcal F_b a(\ov{\widehat \omega^{-\han}\rho_x})
\mathcal F_b^{-1}=a
(\ov{
\mathcal F \widehat \omega^{-\han}\rho_x})$,
it follows that
$
\mathcal F_b\widehat \Phi(x)\mathcal F_b\f=
 \Phi(x)
 $ for each $x$.
 By $\mathcal F \widehat \omega \mathcal F\f=\omega$ it also follows that
$\mathcal F_b\widehat {\hf}\mathcal F_b\f
=\hf  $.
By a simple limiting argument
we can complete the proof.
\qed

We give a remark on the relationship between $H$ and the standard Nelson model $H_N$ introduced in \cite{nel64}.
Namely
\eq{99}
H_N=\lp\otimes 1+1\otimes\hf +\g\Phi_N,\en
where $\d \Phi_N=\int^\oplus_\BR \Phi_N(x) dx$ and
$$\Phi_N(x)=\frac{1}{\sqrt 2}\int
\lk
\frac{\vp(k)}{\sqrt{\omega(k)}}e^{-ikx}\add(k)
+
\frac{\vp(k)}{\sqrt{\omega(k)}}e^{+ikx}a (k)
\rk
dk.
$$
Let $\vm (x)\equiv m^2$ be a nonnegative constant.
Thus the generalized eigenfunction is $\Psi(k,x)=e^{ikx}$ and
$\rho_x=\check\chi(\cdot-x)$.
Then  $H$ covers $H_N$.

\subsection{Klein-Gordon equation on  pseudo Riemannian manifold}
In this subsection we give an example
of a Klein-Gordon equation defined on a pseudo Riemannian manifold $\ms M$ such that a short range potential $\vm(x)=\mathcal O(\ab{x}^{-\beta-2})$ appears, where $\ab{x}=\sqrt{1+|x|^2}$. See \cite{ful96} for details.

Let $\underline x=(t,x)=(x_0,x)\in\RR\times\BR$.
Let
$\ms M$ be
the  $4$ dimensional pseudo Riemannian manifold equipped with
the metric tensor:
\eq{ex10}
g(\underline x)=g(x)=\lk
\begin{array}{cccc}
 {e^{-\theta(x)}} & 0&  0&   0\\
 0& {-e^{-\theta(x)}} & 0& 0 \\
 0& 0& {-e^{-\theta(x)}} & 0  \\
 0& 0& 0& {-e^{-\theta(x)}}
 \end{array}
 \rk.
 \en
Note that $g$ depends on $x$ but independent of $t$.
The line element associated with $g$ is given by
$$ds^2=\eb dt\otimes dt -\eb \sum_j dx^j\otimes dx^j .$$
The Klein-Gordon equation on $\ms M$ is
\eq{ex11}
\square _g\phi+m^2\phi=0,
\en
where
the d'Alembertian operator is defined by $$\square _g=e^{\theta(x)}\partial_t^2 -e^{2\theta(x)}
\sum_j \partial_j e^{-\theta(x)}\partial_j.$$
Thus the
Klein-Gordon equation \kak{ex11} is
reduced to the equation
\eq{ex12}
\frac{\partial^2\phi}{\partial t^2}=K_0 \phi ,\en where
$$K_0= e^{\theta(x)}
\sum_j
\partial_j
\eb \partial_j -\eb m^2.$$
The operator $K_0\lceil_{C_0^\infty(\BR)}$ is symmetric on the weighted $L^2$ space
$L^2(\BR;\eb dx)$.
Now we transform the operator $K_0$
to the one on $\LR$.
In order to do that,  the unitary map $U_0:L^2(\BR;\eb dx)
\rightarrow \LR$ is introduced by
$U_0f(x)=e^{-(\han)\theta(x)}f(x)$.
\bl{ex16}
There exist functions $\theta$ and
$v$ such that
$U_0K_0U_0^{-1}=\Delta-v$,
   $v(x)=\mathcal O(\ab{x}^{-\beta-2})$ for $\beta\geq 0$, and
   $-\Delta +v$ has no non-positive eigenvalues.
\el
Hence the Klein-Gordon equation
\kak{ex12} is transformed
to the equation
\eq{ex12-1}
\frac{\partial^2\phi}{\partial t^2}=
\Delta\phi-v\phi
\en
on $\LR$.
Although the proof of Lemma \ref{ex16} is straightforward, we shall show this statement through a more general scheme in what follows.

Suppose that
 $g=(g_{\mu\nu})$, $\mu,\nu=0,1,2,3$,
is a metric tensor on $\RR^4$ such that
\bi
\item[(1)]
$g_{\mu\nu}(\underline x)=g_{\mu\nu}(x)$, i.e., it is independent of time $t$,
\item[(2)] $g_{0j}(\underline x)=g_{j0}(\underline x)=0$, $j=1,2,3$,
    \item[(3)]
    $g_{ij}(\underline{x})=-\gamma_{ij}(x)$, where $\gamma=(\gamma_{ij})$ denotes a
    $3$-dimensional Riemannian metric.
\ei
Namely
$$g=\MMM {g_{00}} 0 0 {-\gamma}.$$
Let  $\ms M$ be a pseudo Riemannian manifold equipped with the metric tensor   $g$ satisfying (1)-(3) above.
Then
the line element on $\ms M$ is given by
$$ds^2=g_{00}(x)dt\otimes dt -\sum_{ij}\gamma_{ij}(x) dx^i \otimes dx^j.$$
Let $g^{-1}=(g^{\mu\nu})$ denote the inverse of $g$. In particular $1/g_{00}=g^{00}$. We also denote the inverse of $\gamma$ by $\gamma^{-1}=(\gamma^{ij})$.
The Klein-Gordon equation on the static pseudo Riemannian manifold $\ms M$ is generally given by
\eq{ex1}
\square _g \phi+
(m^2+\rieman {\mathcal R})\phi=0,
\en
where $\rieman$ is a constant,
  ${\mathcal R}$ the scalar curvature of $\ms M$,
  and $\square _g$ is given by
\eq{ex23}
\square _g=\sum_{\mu\nu}
    \frac{1}{\sg}\partial_\mu g^{\mu\nu} \sg \partial_\nu .
    \en
      Let us assume that $g_{00}(x)>0$.
Then
\kak{ex1} is rewritten as
\eq{ex22}
\frac{\partial^2\phi}{\partial t^2}=K\phi,
\en
where
$$K=g_{00}\lk
\frac{1}{\sg}\sum_{ij}
\partial_j
\sg \gamma^{ji}\partial_i-m^2-\rieman {\mathcal R}\rk.$$
The operator
$K\lceil_{C_0^\infty(\BR)}$ is symmetric on $L^2(\BR;\rho(x) dx)$, where
\eq{ex20}
\rho=\frac{\sg}{g_{00}}=g_{00}^{-1/2}\sqrt{|{\rm det} \gamma|}.
\en
Now let us transform the operator $K$ on
$L^2(\BR;\rho(x) dx)$ to the one on $\LR$.
Define the unitary operator
$U:L^2(\BR;\rho(x) dx)\rightarrow \LR$ by
$$ Uf=\rho^\han f.$$
Let $\rho_i=\partial_i \rho$ and $\partial_i\partial_j\rho=\rho_{ij}$ for notational simplicity. Furthermore we set $\alpha^{ij}=g_{00}\gamma^{ij}$ and $\partial_k\alpha^{ij}=\alpha^{ij}_k$.
Since $U^{-1}\partial_j U=\partial_j +\frac{\rho_j}{2\rho}$, we have
as an operator identity
\eq{ex3}
U^{-1}\lk
\sum_{ij} \partial_i g_{00}\gamma^{ij}
 \partial_j \rk
 U=g_{00}\sum_{ij} \gamma^{ij} \partial_i\partial_j +V_1+V_2,
 \en
 where
\begin{eqnarray*}
V_1&=&
\sum_{ij} \lk
\alpha^{ij}_i +\alpha^{ij}
\frac{\rho_i }{\rho}\rk\partial_j,\\
V_2&=& \frac{1}{4}
\sum_{ij}
\lk
2
\alpha^{ij}_{i}
\frac{ \rho_j}{\rho}
 +2\alpha^{ij}
\frac{\rho_{ij}}{\rho}-
\alpha^{ij}
\frac{\rho_i}{\rho}
\frac{\rho_j}{\rho}
\rk.
\end{eqnarray*}
Set
$|{\rm det} g|=G$ and $\partial_i G=G_i$.
Hence
we have
$$V_1=g_{00}\sum_{ij} \lk \gamma^{ij}_i+\frac{G_i}{2G}\rk\partial _j,$$where $\gamma^{ij}_i=\partial_i\gamma^{ij}$,
and directly we can see that
\eq{ex4}
g_{00}\frac{1}{\sg}
\sum_{ij} \partial_i
\sg \gamma^{ij}\partial_j=
V_1+g_{00}\sum_{ij}
\gamma^{ij}\partial_i\partial_j.
\en
Comparing \kak{ex3} with \kak{ex4} we obtain that
\eq{ex5}
U^{-1}\lk
 \sum_{ij} \partial_i g_{00}\gamma^{ij}\partial_j-V_2
 \rk U=
 g_{00}\frac{1}{\sg}\sum_{ij} \partial_i
\sg \gamma^{ij}\partial_j.
\en
Then we proved the lemma below.
\bl{ex14}
It follows that
\eq{ex6}
UKU^{-1}=
 \sum_{ij} \partial_i g_{00}\gamma^{ij}\partial_j-v,
 \en where
 $v=g_{00}(m^2+\rieman {\mathcal R})+V_2$.
 \el
By
Lemma \ref{ex14},
\kak{ex22} is transformed to
the equation:
 \eq{23}
\frac{\partial^2\phi}{\partial t^2}=
\lk
 \sum_{ij} \partial_i g_{00}\gamma^{ij}\partial_j-v\rk
 \phi
 \en
on $\LR$.

{\it Proof of Lemma \ref{ex16}}:
Now we come back to the proof of Lemma \ref{ex16}.
Set
$$g_{\mu\nu}(x)=\lkk
\begin{array}{ll}
e^{-\theta(x)},&\mu=\nu=0,\\
-e^{-\theta(x)},& \mu=\nu=1,2,3,\\
0,&\mu\not=\nu.
\end{array}
\right.
$$
Then
\eq{ex7}
\rho=\frac{\sg}{g_{00}}=e^{-\theta}, \quad \alpha^{ij}=g_{00}\gamma^{ij}=\delta_{ij},
\en
and $UKU^{-1}=\Delta -v$ follows by \kak{ex6},
where,
inserting \kak{ex7} to $v$,
we have
\eq{ex8}
v=e^{-\theta}(m^2+\rieman {\mathcal R})
-\frac{\Delta \theta}{2}
+\frac{|\nabla\theta|^2}{4}.
\en
Taking $\rieman=0$, $m=0$, and $\theta(x)=2a \ab{x}^{-\beta}$,
we obtain
\eq{ex21}
v(x)=a\ab{x}^{-\beta-4}(\beta(\beta-1)|x|^2-3\beta) +a^2\ab{x}^{-2\beta-4}|x|^2.
\en
In the case of $0\leq \beta \leq 1$ and $a<0$, we see
that $v\geq 0$ and $v=\mathcal O(\ab{x}^{-\beta-2})$.
Furthermore $-\Delta+v$ has no non-positive eigenvalues.
In the case of $\beta >1$ and $a>0$, we see that however
 $v\not\geq0$.
We can estimate the number of non-positive eigenvalues of  $-\Delta +v$
by the Lieb-Thirring inequality \cite{lie73}:
\eq{lt}
\# \{\mbox{ eigenvalues of }-\Delta+v\leq  0\}\leq C_{LT}
\int |v_-(x)|^{3/2} dx,\en
 where $v_-$ denotes the negative part of $v$ and $C_{LT}$ is a constant independent of $v$.
This yields that
  $-\Delta+v$ has no non-positive eigenvalues for sufficiently small $a$.
Thus the lemma  holds.
\qed
\section{Functional integrations}
\subsection{Path measures for
particles}
In order to construct a functional integral representation
we introduce a probability measure $P^x$
with reference measure $\mu_{\rm p}$
such that
$(f, e^{-t\lp} g)$ can be expressed as
\eq{k26}
(f, e^{-t\lp} g)=\int
\dm \mathbb E ^x[\ov{f(X_0)} g(X_t)].
\en
We already mention that formally
$\lp$ is given by
\eq{lp-1}
\lp f= -\half \Delta f+\frac{\nabla \grp }{\grp}\nabla f.
\en
Thus $X=(X_t)_{t\in\RR}$ is the solution of the stochastic differential equation
\eq{sde}
dX_t=dB_t+\nabla\log \grp(X_t) dt.
\en
The regularity of ground state $\grp$ is, however, unclear.
So we construct the process $X$ through the Kolmogorov consistency theorem.
Let us set $\lpb=\lp-\is(\lp)$.
\bp{lms}
Suppose that Assumption \ref{80} holds.
Then
there exists a probability space
$(\Omega, \ms B, P^x)$ and
an $\BR$-valued continuous
Markov process $X=(X_t)_{t\in\RR}$ starting at $x$
such that
 for $t_0\leq t_1\leq\cdots\leq t_n$ and
$f_0,f_n\in \HP$ and $f_j\in L^\infty(\BR)$,
$j=1,..n-1$,
\eq{8}
(f_0, e^{-(t_1-t_0)\lpb }f_1\cdots e^{-(t_n-t_{n-1}) \lpb }f_n)_{\HP}=
\ix{
\prod_{j=0}^n
f_j(X_{t_j})}.
\en
\ep
\proof
We show an outline of the proof.
The proof is based on the Kolmogorov consistency
theorem.
For $t_0\leq t_1\leq \cdots\leq t_n$ and
$A_j\in \ms B (\BR)$, $j=0,1,...,n$, where ${\ms B} (\BR)$ denotes the Borel
$\sigma$-field,
let
$$\nu(A_0\times \cdots\times A_n)=
(1_{A_0}, e^{-(t_1-t_0)\lpb }1_{A_1}
\cdots e^{-(t_n-t_{n-1}) \lpb }1_{A_n})_{\HP}.
$$
Thus $\nu$ satisfies the consistency condition
$$
\nu(A_0\times \cdots\times A_n\times
\underbrace{
\BR
\times \cdots \times
\BR}_m)
=\nu(A_0\times \cdots \times A_n).$$
By the Kolmogorov consistency theorem
there exists a measure $\nu_\infty$
on
$(\BR)^{(-\infty, \infty)}$ such that
$$\nu(A_0\times \cdots \times A_n)=
\mathbb E_{\nu_\infty}\left[
\prod_{j=0}^n  1_{A_j}(X_{t_j})\right],
$$
where
$X_t(\omega)=\omega(t)$ for $\omega\in
(\BR)^{(-\infty, \infty)}$ the point evaluation.
We note that by the Feynman-Kac formula
$E_{\nu_\infty}[|X_t-X_s|^{2n}]$ can be expressed in terms of Brownian motion $(B_t)_{t\geq 0}$
on $(W, \ms B_W, P_W)$
as
  $$\mathbb E_{\nu_\infty}
  [|X_t-X_s|^{2n}]=
 \int dx \mathbb E_{P_W}
 ^x\left[
 |B_{t-s}-B_0|^{2n}\grp(B_0) \grp(B_{t-s})
 e^{-\int_0^{t-s}V(B_r)dr}
 \right]e^{(t-s)\is(\lp)}.$$
 By (1) of Assumption \ref{80}
we have
$$\sup_{x\in\BR}\mathbb
E_{P_W}
^x
\left[
e^{-\int_0^{t-s}V(B_r)dr}
\right]<\infty,$$
and
$\mathbb E_{P_W}^x[|B_{t-s}-B_0|^{2n}]=C_{2n}|t-s|^n$ with some constant $C_{2n}$.
Then
it can be shown that
$\mathbb E_{\nu_\infty}[|X_t-X_s|^{2n}]
\leq C |t-s|^n$ with some constant $C$ independent of $s$ and $t$.
Then $X=(X_t)_{t\in \RR}$
has a continuous version
$\tilde X=(\tilde X_t)_{t\in\RR}$.
The image measure of $\nu_\infty$
on
$\Omega=C(\RR;\BR)$  with
respect to $\tilde X$
is denoted by
$P$ and
define\footnote{Let
$\s(\tilde X_0)$ denote the
$\s$-filed
generated by $\tilde X_0$.
For $Z\subset \Omega$,
let
$P(Z|\s(\tilde X_0))=\mathbb E_P[1_Z|\s(\tilde X_0)]$.
Then
$P(Z|\s(\tilde X_0))$ is $\s(\tilde X_0)$-measurable.
Thus
$P(Z|\s(\tilde X_0))$ is a function of $\tilde X_0$, i.e.,
$P(Z|\s(\tilde X_0))=G_Z(\tilde X_0)$ with some $G_Z$.
$P(Z|\tilde X_0=x)$ is defined by $G_Z(\tilde X_0)$ with $\tilde X_0$ replaced by $x$, i.e., $P(Z|\tilde X_0=x)=G_Z(x)$.
}
the measure
\eq{meas1}
P^x(\cdot)=P(\cdot|\tilde X_0=x)
\en
for
$x\in\BR$
on $\Omega$.
Then
\eq{ptk}
(1_{A_0}, e^{-(t_1-t_0)\lpb }1_{A_1}
\cdots e^{-(t_n-t_{n-1}) \lpb }1_{A_n})_{\HP}
=\mathbb E^x
\left[\prod_{j=0}^n 1_{A_j}(\tilde X_{t_j})\right].
\en
Here $\mathbb E^x=\mathbb E_{P^x}$.
By a simple limiting argument, \kak{8} can be proven.
Finally we shall show the Markov property of $\tilde X$.
Let \eq{ptk2}
p_t(x, A)=\lk e^{-t\lpb }1_A\rk(x).
\en
Then
\kak{ptk} is represented as
$$\int \prod_{j=0}^n 1_{A_j}(x_j) \prod_{j=1}^n p_{t_j-t_{j-1}}(x_{j-1}, dx_j)\grp^2(x_0) dx_0.$$
Hence it is enough to show that
$p_t(x,A)$ is a probability transition kernel.
Note that
$e^{-t\lpb }$ is positivity preserving.
Then $0\leq e^{-t\lpb } f\leq 1$ for all
function $f$ such that $0\leq f\leq 1$, and $e^{-t\lpb }1=1$ follow.
Then it satisfies that
\bi
\item[(a)]
 $p_t(x,\cdot)$ is the probability measure on $\BR$ with $p_t(x, \BR)=1$,
\item[(b)]
 $p_0(x,A)=1_A(x)$,
 \item[(c)]
 $\int p_s(y,A) p_t(x,dy)=p_{t+s}(x,A)$.
\ei
Hence $p_t(x,A)$ is a probability transition kernel.
Then
the process $\tilde X$ constructed above is Markov under the measure $P^x$.
\qed
By \kak{8}
it can be seen that $X$
is invariant
with respect to
any time shift,
namely
$$\ix{
\prod_{j=0}^n
f_j(X_{t_j})}=
\ix{
\prod_{j=0}^n
f_j(X_{s+t_j})}$$
for any $s\in\RR$.
The time reversal property   also holds:
$$\ix{
\prod_{j=0}^n
f_j(X_{t_j})}=
\ix{
\prod_{j=0}^n
f_j(X_{-t_j})}.$$
Moreover
 $X_t$ and $X_{-s}$ for $-s\leq 0\leq t$ are independent, since
$$\EE^x[X_{-s}X_t]=
\EE^x[X_{-s}\EE^x[X_t|\ms B_{[-s,0]}]]
=\EE^x[X_{-s}\EE^{X_0}[X_t]]
=
\EE^x[X_{-s}] \EE^x[X_t],
$$
where $\ms B_{[a,b]}=\s(X_r,a\leq r\leq b)$.

\subsection{Building of quantum fields
and semigroups}
The free Hamiltonian $\hf$ can be regarded as
the infinite dimensional version of
the harmonic oscillator
$H_{\rm osc}=\half p^2+
\half x^2-\half$.
The process associated with $H_{\rm osc}$
is the Ornstein-Uhlenbeck process
$(q_t)_{t\in\RR}$, and hence
$$\int dx \Psi(x)^2  \EE^x[q_t q_s]=
(x \Psi, e^{-(t-s)H_{\rm osc}}x\Psi)=e^{-|t-s|},$$
 where $\Psi(x)=\pi^{-1/4}e^{-x^2/2}$ is the ground state of $H_{\rm osc}$.
There exists an infinite dimensional version
of $q=(q_t)_{t\in\RR}$.

Let $d=1,2,...$
denote the dimension.
Let $\Phi_d(f)$ be the Gaussian random process
indexed by real-valued $f\in
L^2(\RR^d)$ on
some probability space
$({\ms Q}_d,\mu_d)$ with
mean zero  and the covariance
given by
$$\int_{{\ms Q}_d} \Phi_d(f)\Phi_d(g) d \mu_d=\half
(\hat f, \hat g)_{L^2(\RR^d)}.
$$
The set of the linear hull
of functions of the form
$:\Phi_d(f_1)\cdots \Phi_d(f_n):$ is dense
in
$L^2({\ms Q}_d)$, where $:Z:$ denotes the Wick product of $Z$ inductively defined by
$:\Phi_d(f):=\Phi_d(f)$ and
\eqn
&&
:\Phi_d(f)\Phi_d(f_1)\cdots \Phi_d(f_n):\\
&&
=
:\Phi_d(f_1)\cdots \Phi_d(f_n):
-\half\sum_{j=1}^n
(\bar f,f_j)
:\Phi_d(f_1)\cdots\widehat{\Phi_d(f_j)}\cdots \Phi_d(f_n):,
\enn
where $\widehat{\Phi_d(f_j)}$
denotes neglecting
${\Phi_d(f_j)}$.
Note that
$$(:\Phi_d(f_1)\cdots \Phi_d(f_n):,
:\Phi_d(\rho_1)\cdots \Phi_d(\rho_m):)=\delta_{nm}
\frac{1}{2^n}
\sum_{\sigma\in \ms \mathfrak{G}_n}
(f_1,\rho_{\s(1)})\cdots
(f_n,\rho_{\s(n)}).
$$
For Hilbert spaces $A$ and $B$, let $${\rm \iso }(A,B)=\{T:A\rightarrow B|\|T\|_{A\rightarrow B}\leq 1\}$$ be the set of contarctions from $A$ to $B$, and
$$\HH(A,B)=\{T\in {\rm \iso }(A,B)|T\mbox { is isometry}\}.$$
The second quantization
$\Gamma$ is
a functor:
$$
\Gamma: {\rm \iso }(L^2(\RR^d),L^2(\RR^{d'}))
\rightarrow
{\rm \iso }(L^2(\ms Q_d), L^2(\ms Q_{d'}))$$
and
$$
\Gamma:
\HH
(L^2(\RR^d),L^2(\RR^{d'}))
\rightarrow
\HH
(L^2(\ms Q_d), L^2(\ms Q_{d'})),$$
and it is
defined by $\Gamma(T)1_{L^2(\ms Q_d)}=1_{L^2(\ms Q_{d'})}$ and
\eq{75-1}
\Gamma(T):
\Phi_d(f_1)\cdots \Phi_d(f_n):=
:\Phi_{d'}(Tf_1)\cdots \Phi_{d'}(Tf_n):.
\en
It satisfies the semigroup property:
\eq{semi}
\Gamma(T)\Gamma(S)=\Gamma(TS),
\en
when $S\in{\rm \iso }(L^2(\RR^d),L^2(\RR^{d'}))$ and
$T\in {\rm \iso }(L^2(\RR^{d'}),
L^2(\RR^{d''}))$.
Contraction operator $\Gamma(T) $ depends on $d$ and $d'$,
we do not, however, distinguish them,
and simply write~$\Gamma(T)$.
 $\Gamma(e^{-itK})$ for a self-adjoin operator $K$ in $L^2(\RR^d)$
 is one parameter unitary group on $L^2(\ms Q_d)$.
 Then
its generator is denoted by $d\Gamma(K)$, namely $\Gamma(e^{-itK})=e^{-itd\Gamma(K)}$.

Let $h\geq 0$
be a Borel measurable function on
$\RR^d$.
Define the family of
isometries
$ j_{d,h}(t)\in\HH(
L^2(\RR^d),L^2(\RR^{d+1}))$,
$t\in\RR$, by
\eq{50}
\widehat{j_{d,h}(t)f}=
\frac{e^{-itk_{d+1}}}{\sqrt\pi}
\lk
\frac{h(k)}{h(k)^2+|k_{d+1}|^2}
\rk^{\han} \hat f(k),\quad k\in \RR^d, \quad
k_{d+1}\in\RR.
\en
It satisfies that
\eq{51}
j_{d,h}(s)^\ast
j_{d,h}(t)=e^{-|t-s|h(-i\nabla)}.
\en
For a given Borel measurable nonnegative functions
$h_1$ on $\BR$,
$h_2$ on $\RR^4$,
$h_3$ on $\RR^5....$,
we have a sequence
\eq{52}
\LR
\stackrel{j_{3,h_1}(t)}
{\longrightarrow}
L^2(\RR^4)
\stackrel{j_{4,h_2}(t)}
{\longrightarrow}
L^2(\RR^5)
\stackrel{j_{5,h_3}(t)}
{\longrightarrow}\cdots.
\en
Each isometry in \kak{52} satisfies \kak{51}.
Define $J_{d,h}(t)\in\HH
(L^2({\ms Q}_d),
 L^2({\ms Q}_{d+1}))$
by the second quantization of
$j_{d,h}(t)\in\HH(L^2(\RR^d),L^2(\RR^{d+1}))$, namely $J_{d,h}(t)=\Gamma(j_{d,h}(t))$.
Hence it follows that
\eq{54}
J_{d,h}(s)^\ast  J_{d,h}(t)=
\Gamma(e^{-|t-s|h(-i\nabla)}).
\en
Sequence \kak{52} is inherited
on $L^2({\ms Q}_d)$ as
\eq{55-1}
L^2({\ms Q}_3)
\stackrel{J_{3,h_1}(t)}
{\longrightarrow}
L^2({\ms Q}_4)
\stackrel{J_{4,h_2}(t)}
{\longrightarrow}
L^2({\ms Q}_5)
\stackrel{J_{5,h_3}(t)}
{\longrightarrow}\cdots.
\en
Let
$h$ and $f$ be
Borel measurable
nonnegative functions on $\RR^d$.
The crucial property  is
the intertwining property given
by
\eq{55}
\Gamma(e^{-t(h(-i\nabla)\otimes 1)})
J_{d,f}(s)=
J_{d,f}(s)\Gamma(e^{-th(-i\nabla)}).
\en
Here $h(-i\nabla)\otimes 1=h(-i\nabla)\otimes 1_{L^2(\RR)}$ is an operator on
$L^2(\RR^{d+1})$
under the identification
$L^2(\RR^{d+1})
\cong
L^2(\RR^{d})\otimes L^2(\RR)$.
\bp{62}
Let $h_j$, $j=1,...,N$,
 be Borel measurable nonnegative functions on $\BR$.
 Let $H_j=d\Gamma(h_j(-i\nabla))$.
 Then
\eq{56}
\lk
\Psi, \prod_{i=1}^N
 e^{-t_iH_i}
 \Phi\rk
 _{L^2({\ms Q}_3)}
 =
 \lk
 \prod_{i=N}^1
  J_{i+2,h_i^{\rm ex}}(0)
 \Psi,
 \prod_{i=N}^1
 J_{i+2,h_i^{\rm ex}}(t_i)
 \Phi\rk
 _{L^2({\ms Q}_{N+3})}.
\en
Here $\prod_{i=1}^N T_i=T_1\cdots T_N$ and $\prod_{i=N}^1 T_i=T_N\cdots T_1$ and
$h_i^{\rm ex}$
is an extension of $h$ to the nonnegative function  on $L^2(\RR^{2+i})$ defined by
$h_i^{\rm ex}
({\bf k},k_4,...,k_{2+i})
=
h_i({\bf k})$ for ${\bf k}\in\BR$.
\ep
In order to construct
a functional integral representation
of the semigroup $e^{-tH}$
we take the Schr\"odinger representation
instead of
the Fock representation.
In addition we need the Euclidean field.
We set
\eq{77}
\begin{array}{lll}
\ms Q=
{\ms Q}_3,&
 \mu=\mu_3,  &  j_t= j_{3,\omega}(t), \\
 \ms Q_E={\ms Q}_4,
&
\mu_E=\mu_4, &  \xi_t= j_{4,I}(t),
\end{array}
\en
where $I$ denotes the identity operator on $L^2(\RR^4)$.
It is well know that there exists
an isomorphism between
$\ms F$ and $L^2({\ms Q})$.
By this isomorphism  we can identify
as
$
\Omega_\ms F\cong 1$,
$
\hf\cong d\Gamma(\omega(-i\nabla))
$
and
$\Phi(x)\cong \phi(\tvp(x))$,
where
\eq{41}
\tvp(\cdot,x)=
\lk
\frac{\vp(\cdot)}{\sqrt{\omega(\cdot)}}
\ov{\Psi(\cdot,x)}\rk^\vee.
\en
Note that in the
Schr\"odinger representation the test function is taken
in the position representation
while the momentum representation
is used in the Fock representation.
\begin{definition}
{\bf (The Nelson model in Schr\"odinger representation)}

In the Schr\"odinger representation
the Nelson Hamiltonian
is defined by
\eq{78}
\lpb\otimes 1+1\otimes d\Gamma(\omega(-i\nabla))
+\g\int_\BR^\oplus
\phi(\tvp(x))dx
\en
on $\HP\otimes \QQ$.
Here we identify $\HP\otimes \QQ$ as
$\int_\BR^\oplus L^2(\ms Q)d\mu_{\rm p}$.
\end{definition}
In what follows we write
 \kak{78} as $H$,
$d\Gamma(\omega(-i\nabla))$ as
$\hf$ and
$\HP\otimes \QQ$ as $\hhh$.

The operator
 $d\Gamma(I)$ is called the number operator.
 The number operator
on $\QQ$ (resp $\QQQ$)
is denoted by $N$ (resp $N_E$).
We define
the specific families of isometries
$J_t\in{\rm \iso }_0(\QQ,\QQQ)
$
and
$\Xi_t\in {\rm \iso }_0(\QQQ,L^2({\ms Q}_5))$ by
\eq{57}
\begin{array}{l}
J_t=\Gamma(j_t) = J_{3,\omega}(t),\\
\Xi_t=\Gamma(\xi_t)=J_{4,I}(t)
\end{array}
\en
for $t\in\RR$.
Thus it follows that
\eq{59}
\begin{array}{l}
J_s^\ast J_t=e^{-|t-s|\hf}\\
\Xi_s^\ast \Xi_t=e^{-|t-s|N_E}.
\end{array}
\en
Moreover we have
\eq{61}
e^{-\beta N_E}J_s=J_se^{-\beta N},\quad \beta\geq 0,
\en
by the intertwining property \kak{55}.
\begin{example}
From Proposition \ref{62} it follows that
\eq{63-1}
(\Psi, e^{-\beta N}e^{-t\hf}\Phi)_{\QQ}=
(\Xi_0 J_0\Psi, \Xi_\beta J_t \Phi)_{L^2({\ms Q}_5)}.
\en
\end{example}

\subsection{Functional integral representations}
Combining the functional integral representations of both $e^{-t\lpb }$ and $e^{-t\hf}$ stated in the previous sections,  we can construct the functional integral representation of $e^{-tH}$

Let
$$\phi_s(f)=\Phi_4(j_s f),\quad
s\in\RR.$$
It is the Gaussian random process
 indexed by real-valued functions $f\in\LR$
 such that the mean is zero
 and the covariance is given by
\eq{12}
\int _{\ms Q}
\phi_s(f)\phi_t(g)d\mu_E
=\int_\BR \ov{\hat f(k)} \hat g(k) e^{-|t-s|\omega(k)}dk.
\en
Thus $(\phi_s(f))_{s\in\RR}$
denotes the infinite dimensional version of the Ornstein-Uhlenbeck process.
We note that
$J_s:\phi(f_1)\cdots\phi(f_n):
=
:\phi_s(f_1)\cdots\phi_s(f_n):$
and
$
J_s1_{\QQ}=1_\QQQ$.
Combining the process
$X_t$ in \kak{8} and $J_t$ in \kak{57}
we obtain  the theorem below.
\bt{fir}
Suppose Assumptions \ref{80},
 \ref{ge} and \ref{sa}.
Let $F,G\in\HP\otimes\QQ$. Then
\eq{14}
(F,e^{-tH}G)=
\int \dm \EE^x\left[
\lk
J_0F(X_0), e^{-\g\int_0^t\phi_s(\tvp(X_s))ds}J_tG(X_t)
\rk
_\QQQ
\right]
\en
\et
\proof
By
the Trotter product formula
$$e^{-tH}=s-\lim_{n\rightarrow\infty}
\lk e^{-(t/n)\lpb }e^{-(t/n)
\g \phi(\tvp(x))}
e^{-(t/n)\hf}\rk^n,$$
 the factorization formula \kak{59}, Markov property of $E_t=J_t J_t^\ast$ and \kak{8}, we have
\eq{14-1}
(F,e^{-tH}G)=
\lim_{n\rightarrow \infty}
\int \dm \EE^x\left[
\lk
J_0F(X_0), e^{-\g \sum_{j=0}^n
\frac{t}{n}
\phi_{tj/n}(\tvp(X_{tj/n}))}J_tG(X_t)
\rk
_\QQQ
\right].
\en
Note that
$s\mapsto \tilde \chi(\cdot,X_s)$ is strongly continuous as the map $\RR\rightarrow \LR$ almost surely.
Hence
$s\mapsto \phi_{s}(\tvp(X_{s}))$
is strongly continuous as the map
$\RR\rightarrow \QQQ$.
By a simple limiting argument
we complete the proof.
\qed
Next let
$$\phi_{s,t}(f)=\Phi_5(\xi_t j_s f),\quad
s,t\in\RR.$$
It is also the Gaussian random process
indexed by real-valued functions $f\in\LR$
with  mean zero  and
the covariance given
by
\eq{45}
\int_{\ms Q_E}
\phi_{s,t}(f)
\phi_{s',t'}(g)d\mu_E =
\half\int
\ov{
\hat f(k)}\hat g(k) e^{-|s-s'|\omega(k)}
e^{-|t-t'|}dk.
\en
We see that
$\Xi_t:\phi_{s_1}(f_1)\cdots\phi_{s_n}(f_n):=
\phi_{s_1,t}(f_1)\cdots\phi_{s_n,t}(f_n):$ and
$\Xi_t 1_\QQQ=1_{L^2(\ms Q_5)}$.
Then we have the theorem.
\bt{fir2}
Suppose Assumptions \ref{80},
 \ref{ge} and \ref{sa}.
Let $F,G\in\hhh$. Then
\eqnn
&&
\hspace{-0.5cm}
\lk
F,e^{-sH} e^{-\beta N} e^{-tH}G
\rk
\non \\
&&\hspace{-0.5cm}
=
\int \dm
\EE^x\left[
\lk
\Xi_0 J_0F(X_0),
e^{-\g \int_0^s\phi_{r,0}
(\tvp(X_r))dr}
e^{-\g \int_s^{s+t}
\phi_{r,\beta}
(\tvp(X_r))dr}\Xi_\beta J_tG(X_t)
\rk
_{L^2({\ms Q}_5)}
\right]\non\\
&&
\ennn
\et
\proof
Throughout this proof we set $\prod_{j=0}^n T_j=T_0T_1\cdots T_n$.

Simply  we put  $\g\phi(\tvp(x))=\phi$.
By the Trotter product formula we have
\eqn
&&
\lk
F,e^{-sH} e^{-\beta N} e^{-tH}G
\rk\\
&&
=\limn \limm
\lk
F,
\lk
e^{-\frac{s}{n}\lpb }e^{-\frac{s}{n}\phi}
e^{-\frac{s}{n}\hf}\rk^n
e^{-\beta N}
\lk
e^{-\frac{t}{m}\lpb }e^{-\frac{t}{m}
\phi}e^{-\frac{t}{m}\hf}\rk^m
G
\rk.
\enn
Inserting
$e^{-|T-S|\hf}=J_T^\ast J_S$ we have
\eqn
&&
\hspace{-0.5cm}=
\lk
F, J_0^\ast
\prod_{i=0}^{n-1}
\lk
J_{\frac{si}{n}}
e^{-\frac{s}{n}\lpb }
e^{-\frac{s}{n}\phi}J_{\frac{si}{n}}^\ast
\rk
J_s e^{-\beta N} J_s^\ast\right.\\
&&\left.\hspace{2cm}
\prod_{i=0}^{m-1}\lk J_{s+\frac{ti}{m}}e^{-\frac{t}{m}\lpb }
e^{-\frac{t}{m}\phi}J_{s+\frac{ti}{m}}^\ast
\rk J_{s+t}G\rk.
\enn
Let $E_T=J_T J_T^\ast $. $E_T$ is the family of projection on $\QQQ$.
Since $J_T^\ast e^\phi J_T=E_T e^{\phi_T}E_T$ and by the intertwining  property
$J_s e^{-\beta N}J_s^\ast=J_s^\ast J_s e^{-\beta N_E}=E_s \Xi_0^\ast \Xi_\beta$, we have
\eqn
&&=
\lk
F, J_0^\ast
\prod_{i=0}^{n-1}
\lk
E_{\frac{si}{n}}
e^{-\frac{s}{n}\lpb }
e^{-\frac{s}{n}\phi_{\frac{si}{n}}}
E_{\frac{si}{n}}
\rk
E_s \Xi_0^\ast \Xi_\beta \right.\\
&&\hspace{5cm}
\left.
\prod_{i=0}^{m-1}\lk
E_{s+\frac{ti}{m}}
e^{-\frac{t}{m}\lpb }
e^{-\frac{t}{m}\phi_{s+\frac{ti}{m}}}
E_{s+\frac{ti}{m}}
\rk J_{s+t}G\rk,
\enn
where $\phi_T=\g\phi_T(\widetilde\chi(x))$.
By the Markov property of $E_s$ we can neglect all $E_s$, then we have
\eqn
&&=
\lk
F, J_0^\ast
\prod_{i=0}^{n-1}
\lk
e^{-\frac{s}{n}\lpb }
e^{-\frac{s}{n}\phi_{\frac{si}{n}}}
\rk
\Xi_0^\ast \Xi_\beta \right.
\left.
\prod_{i=0}^{m-1}\lk
e^{-\frac{t}{m}\lpb }
e^{-\frac{t}{m}\phi_{s+\frac{ti}{m}}}
\rk J_{s+t}G\rk.
\enn
Again we use the fact
$\Xi_\beta e^{\phi_s}\Xi_\beta^\ast
=
E_\beta ^\Xi
e^{\phi_{s,\beta}}E_\beta ^\Xi$, where $E_\beta^\Xi=\Xi_\beta \Xi_\beta ^\ast$ denotes the projection on $L^2(\ms Q_5)$.
Hence  we have
\eqn
&&=
\lk
\Xi_0 J_0F,
E_0^\Xi
\prod_{i=0}^{n-1}
\lk
e^{-\frac{s}{n}\lpb }
e^{-\frac{s}{n}\phi_{\frac{si}{n},0}}
\rk
E_0^\Xi
\right.\\
&&\left.\hspace{2cm}
E_\beta^\Xi
\prod_{i=0}^{m-1}\lk
e^{-\frac{t}{m}\lpb }
e^{-\frac{t}{m}\phi_{s+\frac{ti}{m},\beta}}
\rk
E_\beta ^\Xi\Xi_\beta
J_{s+t}G\rk.
\enn
Since by the Markov property of $E_s^\Xi$ we can neglect $E_0^\Xi$ and $E_\beta^\Xi$, we can obtain
\eqn
&&=
\lk
\Xi_0 J_0F,
\prod_{i=0}^{n-1}
\lk
e^{-\frac{s}{n}\lpb }
e^{-\frac{s}{n}\phi_{\frac{si}{n},0}}
\rk
\prod_{i=0}^{m-1}\lk
e^{-\frac{t}{m}\lpb }
e^{-\frac{t}{m}\phi_{s+\frac{ti}{m},\beta}}
\rk
\Xi_\beta
J_{s+t}G\rk,
\enn
where $\phi_{S,T}=\phi_{S,T}(\widetilde X(x))$.
By \kak{8} and a limiting argument, we can prove the theorem.
\qed

\section{Infrared divergence and absence of ground states}
\subsection{Abstract theory of the absence of ground states}
In this section
we assume
Assumptions \ref{80}, \ref{ge}  and \ref{sa}.
By the functional integral
representation obtained in Theorem \ref{fir}, we can see that
$$(F, e^{-tH}G)>0$$
for any $F\geq 0$ and $G \geq 0$ but
$F\neq$ and $G\neq0$.
Thus $e^{-tH}$ is positivity improving.
Then
whenever a ground state $\gr$ of
$H$ exits, $\gr>0$ by
the Perron-Frobenius Theorem.
In particular
the ground state is unique
if it exists.
Now we introduce a sequence
  approaching to the ground state.
Let
$
1=1_{\HP}\otimes 1_{\QQ}
$
and
\eq{16}
\grt=\|e^{-TH }1\|^{-1}
e^{-TH}1,\quad T>0.
\en
Define
\eq{15}
\gamma(T)=(1,\grt)^2,\quad T>0.
\en
If $H$ has a ground state, then $\grt$ converges
to $\gr$ strongly as $T\rightarrow \infty$.
We can have a criteria on the existence and non-existence of the
ground state.
\bp{criteria}
(1) When $\lim_{T\rightarrow \infty}\gamma(T)=a>0$, $H$ has a ground state.
(2) When
$\lim_{T\rightarrow \infty}\gamma(T)=0$, $H$ has no  ground state.
\ep
Note that
$$\gamma(T)=
\frac{(1,e^{-TH}1)^2}
{\|e^{-TH}1\|^2}.$$
Since
  $\phi_s(g)$ is a Gaussian random process,
by means of the functional integral
representation \kak{14}, we can see that
\begin{eqnarray*}
(1, e^{-TH}1)
&=&
\ix{
e^{(\g^2/2)\lk
\int_0^T\phi_s(\tvp(X_s))ds,
\int_0^T\phi_t(\tvp(X_t))dt
\rk}}\\
&=&
\ix{e^{(\g^2/2)
\int_0^Tds\int_0^Tdt
W(X_s,X_t,|s-t|)}},
\end{eqnarray*}
where
\eq{16-1}
W(X,Y,|t|)=\int\frac{\vp(k)^2}{2\omega(k)}
\ov{\Psi(k,X)}{\Psi(k,Y)}e^{-|t|\omega}dk.
\en
Note that
\eq{po}
\int_0^Tds\int_0^Tdt
W(X_s,X_t,|s-t|)
>0
\en
follows,  since
the left hand side is expressed as
$(
\int_0^T \phi_s(\tvp(X_s))ds,
\int_0^T \phi_t(\tvp(X_t))dt)
$.
While
\begin{eqnarray*}
\|e^{-TH}1\|^2
&=&
\ix{e^{(\g^2/2)\int_0^{2T}ds\int_0^{2T}dt
W(X_s,X_t,|s-t|)}}\\
&=&
\ix{e^{(\g^2/2)\int_{-T}^T
ds\int_{-T}^T dt
W(X_s,X_t,|s-t|)}}
\end{eqnarray*}
by the shift invariance of $X_t$.
Then $\gamma(T)$ can be expressed as
\eq{17}
\gamma(T)=
\frac{\lk
\ix{e^{(\g^2/2)\int_0^Tds\int_0^Tdt
W(X_s,X_t,|s-t|)}}\rk^2}
{\ix{e^{(\g^2/2)\WTT}}}
.
\en
Let $\mu_T$ be the probability measure on $(\BR\times \Omega, \ms B(\BR)\times \ms B)$ defined by for $A\times B\in \ms B(\BR)\times\ms B$,
\eq{17-1}
\mu_T(A\times B)=\frac{1}{Z_T}\ix
{1_{A\times B} e^{(\g^2/2) \WTT}},
\en
where $Z_T$ denotes the normalizing constant
such that $\mu_T$ becomes a probability measure.
\bl{upper}
Integral $\WT$ is real and
it follows that
\eq{18}
\gamma(T)\leq \EE_{\mu_T}
\left[
e^{-\g^2\WT}
\right]
\en
\el
\proof
The numerator of \kak{17}
can be estimated by the Schwartz inequality and
the time shift of $X$
as
\begin{eqnarray*}
&&
\lk
\ix{e^{(\g^2/2)\int_0^Tds\int_0^Tdt
W}}\rk^2\\
&&
\leq
\int \dm
\lk
\EE ^x\left[
e^{(\g^2/2)\int_0^Tds\int_0^T dt W}
\right]\rk
\lk
\EE ^x\left[
e^{(\g^2/2)\int_0^Tds\int_0^T dt W}
\right]\rk
\\
&&=
\int \dm
\lk
\EE ^x\left[
e^{(\g^2/2)\int_0^T ds \int_0^T dt W
}\right]\rk
\lk\EE ^x\left[
e^{(\g^2/2)\int_{-T}^0
ds
\int_{-T}^0 dt W}
\right]\rk.
\end{eqnarray*}
Since $X_t$ and $X_s$ for $s\leq 0\leq t$ are independent,
we have
$$
=
\int \dm
\EE ^x\left[
e^{(\g^2/2)
\lk
\int_0^Tds\int_0^T dt W
+\int_{-T}^0
ds\int_{-T}^0dt W\rk}
\right].
$$
Moreover
from
$\int_{-T}^0
\int_{-T}^0
+\int_0^T
\int_0^T =
\int_{-T}^T
\int_{-T}^T
-2\int_{-T}^0
\int_0^T $ and \kak{po},
it follows that integral
$\WT$ is real and
$$=
\ix{e^{-\g^2\wt+(\g^2/2)\wtt}}.$$
Then the lemma follows.
\qed
We can compute $W$ explicitly.
Note that
the
operator
$e^{-|t|\sqrt{-\Delta +m^2}}$ has the integral kernel
$$
e^{-|t|\sqrt{-\Delta +m^2}}(X,Y)=2\lk\frac{m}{2\pi}\rk^{(d+1)/2}
\frac{|t|}{(|X-Y|^2+|t|^2)^{(d+1)/4}}
K_{\frac{d+1}{2}}(m\sqrt{|X-Y|^2+t^2}),$$
where
$K_\nu$ denotes the modified Bessel function of the third kind.
In particular
in the case of $d=3$ and $m=0$
we have
$$
e^{-|t|\sqrt{-\Delta}}(X,Y)=
\frac{1}{\pi^2}
\frac{|t|}{(|X-Y|^2+|t|^2)^2}\quad (d=3).$$
Then
\begin{eqnarray*}
W(x,y,|T|)
&=&
\half \int_T^\infty
 d|t|\lk
  {\Psi_x}\vp, e^{-|t|\omega}{\Psi_y} \vp
  \rk\\
  &=&
  \frac{1}{4\pi^2}
 \int dX\int dY
\frac{ \ov{({\Psi_x}\vp)^\vee}(X)
 ({\Psi_y} \vp)^\vee(Y)
}{|X-Y|^2+|T|^2}.
\end{eqnarray*}
We are in the position to state
the main theorem.
This is an abstract version of \cite{lms}. \bt{hiroshima4}
Let
$A_T=\BR\times\{\tau\in \Omega||X_s(\tau)|\leq T^\lambda, |s|\leq T\}$
for some $\la$ such that
\eq{assalp}
\frac{1}{\AAA +1}<\la<1,
\en
where $\AAA $ is the positive constant given in Assumption \ref{80}.
Suppose that there exists
$\varrho(T)$ independent of $\tau\in\Omega$
such that
\eq{63}
1_{A_T}
\int_{-T}^0 ds \int_0^T dt
 \int dX\int dY
\frac{ {(\ov{\Psi_{X_s}}\vp)^\vee}(X)
 ({\Psi_{X_t}} \vp)^\vee(Y)
}
{|X-Y|^2+|s-t|^2}\geq \varrho(T)
\en
and
$\lim_{T\rightarrow \infty}
\varrho(T)=\infty$.
Then there is no ground states of $H$.
\et
\proof
By Lemma \ref{upper}
it is enough to show that
\bi
\item[(1)]
$\d \lim_{T\rightarrow \infty}
\EE_{\mu_T}\left[
1_{A_T}e^{-\g^2\WT}\right]=0$,
\item[(2)]
$\d \lim_{T\rightarrow \infty}
\EE_{\mu_T}\left[
1_{A_T^c}
e^{-\g^2\WT}\right]=0.$
\ei
(1) follows from assumption \kak{63}.
We shall prove (2).
Note that
\eq{65}
\int_{-T}^0ds\int_0^T dt
e^{-|t-s|\omega}
=
\frac{1}{\omega^2}
\lk
e^{-T\omega}-1
\rk^2
\en
and
\eq{66}
\int_{-T}^Tds
\int_{-T}^T dt
e^{-|t-s|\omega}
=
\frac{2}{\omega^2}
\lk
e^{-2T\omega}-1+2T\omega\rk
.
\en
Then
$$\left|
\WT
\right|\leq \frac{T}{2}
 \|\vp/\omega\|^2$$ and
\begin{eqnarray}
&&\EE_{\mu_T}
\left[
1_{A_T^c}
e^{-\g^2\WT}
\right]\non
\\
&&
\leq
e^{\g^2(T/2)\|\vp/\omega\|^2}
\frac{
\ix{1_{A_T^c}e^{(\g^2/2)\wtt}}}
{\ix{e^{(\g^2/2)\wtt}}}\non
\\
&&
\label{67}
\leq
e^{\g^2(T/2)\|\vp/\omega\|^2}
\frac{
\lk
\ix{e^{\g^2\wtt}}\rk^\han}
{\ix{e^{(\g^2/2)\wtt}}}
\ix{1_{A_T^c}}.
\end{eqnarray}
Moreover by \kak{66},
there exists a constant $\delta>0$
such that
\eq{25-1}
-T\delta \|\vp/\omega\|^2\leq
\WTT\leq T\delta \|\vp/\omega\|^2.
\en
Then we have
\eq{26}
\frac{
\lk
\ix{e^{\g^2\WTT}}\rk^\han}
{\ix{e^{(\g^2/2)\WTT}}}\leq
e^{\g^2\delta T\|\vp/\omega\|^2}.
\en
The crucial part
is to show
that
there exists
an at most polynomially growth
function $\xi(T)$ such that
\eq{27}
\ix{1_{A_T^c}}\leq \xi(T) \exp\lk
-cT^{\lambda(\AAA +1)}\rk.
\en
This is proven in Lemma \ref{lms2} below. Combining \kak{67}, \kak{26} and \kak{27} we have
\eq{28}
\lim_{T\rightarrow\infty}
\EE_{\mu_T}[1_{A_T^c}]
\leq
\lim_{T\rightarrow\infty}
\xi(T) e^{-cT^{\lambda(\AAA +1)}}
e^{\g^2(\delta+\han)
 T\|\vp/\omega\|^2}=0,
\en
since $\frac{1}{\AAA +1}<
\lambda<1$.
Then (2) follows.
\qed
It remains to show \kak{27}.
\bl{lms2}
\kak{27} holds.
Explicitly
$\lim_{T\rightarrow\infty}
\xi(T)/T^{\frac{1-2\la}{2}}<\infty$.
\el
\proof
Recall  that the external potential is supposed to be
$V(x) >|x|^{2\AAA }$ for sufficiently large $|x|$,  and $V_+\in L_{\rm loc}^1(\BR)$ and $V_-\in L^p(\BR)$ with $p> 3/2$.
Then by \cite{car}, the ground state $\gr$ of $\hp$
exponentially decays.
More explicitly
there exist constants
$C>0$ and $\delta>0$ such that
\eq{car1}
\grp(x)\leq Ce^{-\delta|x|^{\AAA +1}}.
\en
We divide the left hand side of \kak{27}
as
\eq{div}
\int_\BR \mathbb
E^x\left[
\sup_{|s|<T^\la}|X_s|>T^\la
\right]\grp(x)^2 dx
=\int_{|x|<T^\la/2}+
\int_{|x|\geq T^\la/2}=Q_1+Q_2.
\en
Let $D_a(n)=\{aj/2^n|j=0,1,..,2^n\}$ be the set of diadic points.
By \cite[Lemma~1.12]{var}
it follows that
\eq{va11}
\mathbb E^0\left[\sup_{0\leq s\leq a, s\in D_a(n)}|f(X_s)|>b\right]
\leq
\frac{3}{b}\sqrt{(f,f)+a
(\lpb^\han f, \lpb^\han  f)}
\en
for $f\in D(\lpb^\han )$, where $(f,g)=(f,g)_{L^2(\BR;\grp(x)^2dx)}$.
The right-hand side above is uniformly
bounded with respect to $n$, and the indicator function $1_
{\{
\sup_{|s|<a, s\in D_a(n)}|f(|X_s|)|>b\}}$ is monotonously increasing in $n$ and $X_t(\omega)$ is continuous in $t$ for each path $\omega$.
Thus by the monotone convergence theorem, we have \begin{eqnarray*}
\lim_{n\rightarrow\infty}
\mathbb E^0\left[\sup_{0\leq s\leq a, s\in D_a(n)}|f(X_s)|>b\right]
&=&
\mathbb E^0\left[
\lim_{n\rightarrow\infty}
\sup_{0\leq s\leq a, s\in D_a(n)}|f(X_s)|>b\right]\\
&=&
\mathbb E^0\left[
\sup_{0\leq s\leq a}
|f(X_s)|>b\right].
\end{eqnarray*}
Hence
\eq{va1}
\mathbb E^0\left[\sup_{|s|<a}|f(X_s)|>b\right]
\leq
2 \frac{3}{b}\sqrt{(f,f)+a (\lpb^\han f, \lpb^\han  f)}
\en
follows.
We apply \kak{va1} to \kak{div}.
Suppose that
$\ggg \in C^\infty(\BR)$
and
$$\ggg (x)=\lkk
\begin{array}{ll}
|x|,& |x|\geq T^\la,\\
0,& |x|\leq T^\la-1.
\end{array}\right.
$$
Moreover we assume that
\eq{note}
e^{-(\delta/2) |x|^{\AAA +1}}f^2,
\quad
e^{-(\delta/2)
|x|^{\AAA +1}}\partial_\mu f\cdot f,
\quad
e^{-(\delta/2) |x|^{\AAA +1}}\partial_\mu^2
f\cdot f\in\LR,\quad \mu=1,2,3,
\en
and the $L^2$ norm of each terms in \kak{note}
has a upper bound independent of $T$.
By \kak{va1} for $T^\la+b>0$,
\begin{eqnarray}
\mathbb E^0\left[\sup_{|s|<a}|
\ggg (X_s)|>T^\la+
b\right]
&=&
E^0\left[\sup_{|s|<a}|X_s|>T^\la+b\right] \non\\
& \leq&\label{va2}
\frac{6}{T^\la+b}\sqrt{(\ggg ,\ggg )+
a (\ggg , \lpb \ggg )}.
\end{eqnarray}
Let $|x|<T^\la/2$.
Thus we have
\begin{eqnarray*}
&&
\mathbb E^x\left[\sup_{|s|<T}|X_s|>T^\la\right]
=
\mathbb E^0\left[\sup_{|s|<T}|X_s+x|>T^\la\right] \\
&&
\leq
\mathbb E^0\left[\sup_{|s|<T}|X_s|>T^\la-|x|\right]
\leq
\frac{6}{T^\la/2}
\sqrt{(\ggg ,\ggg )+
T  (\ggg , \lpb \ggg )}.
\end{eqnarray*}
We estimate the right-hand side above.
By \kak{car1}
we have
\eq{va3}
(\ggg ,\ggg )=\int \ggg (x)^2\grp(x)^2dx\leq
C^2 e^{-\delta T^{\la(\AAA +1)}}
\int \ggg (x)^2
e^{-\delta |x|^{\AAA +1}}
dx:=
a_1
e^{-\delta T^{\la(\AAA +1)}}.
\en
While
\begin{eqnarray*}
(\ggg , \lpb \ggg )
&=&
-\is(\lp)(\ggg,\ggg)+\int \grp(x)^2 \cdot
\ggg (x) \frac{1}{\grp(x)}
\lk
-\half \Delta +V(x)\rk
\grp(x) f(x) dx\\
&=&
-\is(\lp)(\ggg,\ggg)+
\int \grp(x)^2 f(x)^2 V(x) dx
-\half
\int \grp(x) \ggg (x)\Delta
(\ggg \grp)(x).
\end{eqnarray*}
Then the first term on the right-hand side above
is
\eq{va4}
\int \grp(x)^2 f(x)^2 V(x) dx\leq
C^2e^{-\delta T^{\la(\AAA +1)} }
\int e^{-\delta |x|^{\AAA +1}}
 f(x)^2 |x|^{2\AAA }
  dx:=a_2
  e^{-\delta T^{\la(\AAA +1)} }
\en
  and the second term is
\begin{eqnarray}
&&
\int \grp(x) \ggg (x)\Delta
(\ggg \grp)(x)dx
\non
\\
&&=
\int \grp(x)\cdot
\underbrace{
 \lk
 \ggg (x)^2 \Delta \grp(x)+
2 f(x) \nabla\grp(x) \cdot
\nabla f(x) +\Delta \ggg (x)
\cdot f(x)\grp(x)\rk}_{=G(x)}
 dx\non
 \\
 &&
 \leq
C e^{-(\delta/2)
 T^{\la(\AAA +1)}}
\int
e^{-(\delta/2)
 |x|^{\AAA +1}}
|G(x)|dx=a_3  e^{-(\delta/2)
 T^{\la(\AAA +1)}}
.
\label{va5}
 \end{eqnarray}
 Hence
 \eq{g1}
  Q_1
  \leq
 \frac{12}{T^\la}
 \sqrt{|a_1-\is(\lp)|
 +T\lk
 a_2 +a_3\rk}
 e^{-(\delta/4) T^{\la(\AAA +1)}}.
\en
Moreover
\eq{g2}
 Q_2\leq
C^2
 e^{-\delta T^{\la(\AAA +1)}}
 \int
 e^{-\delta |x|^{\AAA +1}} dx=a_4
 e^{-\delta T^{\la(\AAA +1)}}.
\en
 \kak{g1} and \kak{g2}
 yield that
 \eq{yama}
 \mathbb E_{\mu_T}\left[
 1_{A_T^c}
 \right]
\leq \xi(T) e^{-(\delta/4) T^{\la(\AAA +1)}},\en
where
$\xi(T)=\frac{12}{T^\la}
 \sqrt{|a_1-\is(\lp)|
  +T(a_2 +a_3)}
+a_4$.
This completes the proof.
 \qed

\subsection{Absence of ground state for
short range potentials}
In this subsection
we give an example
for a short range variable mass $\vm$.
We introduce
the assumption below:
\begin{assumption}
\label{shortw}
Let $\vm$ be of the form
$\vm =\kappa w$
with $\kappa>0$,
where
$w : \mathbb{R}^3 \rightarrow \RR$ is bounded,
$-\Delta+w$ has no non-positive eigenvalues,
and
there exist positive constants
$C$, $R$ and $\beta>3$ such that
$|w(x)| \leq C \langle x\rangle^{-\beta}$.
\end{assumption}
Assumption \ref{shortw} yields that
 there exists
a generalized eigenfunction $\Psi_\kappa(k,x)$ satisfying
$(-\Delta +\vm)\Psi_\kappa(k,x)=|k|^2\Psi_\kappa(k,x)$
and the Lippman-Schwinger equation
\eq{LipSch} \Psi_\kappa(k,x)
		= e^{ikx}-\frac{\kappa}{4\pi}
			\int \frac{e^{i|k||x-y|}w(y)}{|x-y|}\Psi_\kappa(k,y)dy
\en
by \cite{Ikebe}.
\bl{ikebe2}
Suppose Assumption \ref{shortw}.
Then
\bi
\item[(1)]
$\Psi_\kappa(k,x)$ is continuous in $x$ for each $k$ but $k\not=0$;
\item[(2)]
the generalized Fourier transformation $\mathcal{F}$ define by \kak{94} with $\Psi_\kappa$ is unitary on $\LR$;
\item[(3)]
 there exist positive constants $\kappa_0>0$ and $C_0>0$ such that,
for any $\kappa \leq \kappa_0$,
\eq{68}
\sup_{k\in\BR}
|e^{ikx}-\Psi_\kappa(k,x)|
\leq \kappa C_0\ab{x}^{-1};
\en
\item[(4)]
$\sup_{x,k}|\Psi_\kappa(k,x)| < \infty$
uniformly for
 sufficiently small $\kappa$.
\ei
In particular
 $\vm$ satisfying Assumption \ref{shortw}
fulfills Assumption \ref{ge}.
\el
\proof
(1) follows from \cite{Ikebe},
and (2) again  from \cite{Ikebe} since there exist no non-positive eigenvalues
for $-\Delta + \kappa w$.
We prove (3).
In general there exists a constant $c$ such that
$$\int_{\RR^n}\frac{1}{|x-y|^a\ab{y}^b}
dy\leq c\frac{1}{\ab{x}^a},$$
if $0<a<n<b$.
Then by the assumption $\beta>3$,
we have
$$\d \int_\BR \frac{1}{|x-y|\ab{y}^\beta}dy\leq
c'\frac{1}{\ab x}$$
with some constant $c'$.
Iterating \eqref{LipSch}, we have
\eq{suzuki}
e^{ikx}-\Psi(k,x)
=
\sum_{n=1}^\infty
\lk
\frac{\kappa}{4\pi}\rk^n
\int\cdots\int
\frac{e^{i|k|\sum_{j=1}^n |y_j-y_{j-1}|}
\prod_{j=1}^nw(y_j)}
{\prod_{j=1}^n |y_j-y_{j-1}|}
dy_1\cdots dy_n
\en
with $y_0=x$.
Note that
$$\int  \frac{|w(y)|}{|x-y|}dy \leq \sup_{y\in\BR}|w(y)\ab{y}^\beta |\int\frac{1}{|x-y|\ab{y}^\beta} dy\leq C\ab{x}^{-1}$$
with some constant $C$.
The right hand side of \kak{suzuki}
 absolutely converges for sufficiently small $\kappa>0$.
By \kak{suzuki}
it follows that
\[ |\Psi_\kappa(k,x) - e^{ikx}|
		\leq \sum_{n=1}^\infty \left( \frac{\kappa C}{4\pi} \right)^n \ab{x}^{-1}
			= \frac{\kappa C}{4\pi - \kappa C}\ab{x}^{-1}.
	\]
This completes (3). (4) is derived from (3). The proof is complete.
\qed
Henceforth, we denote $\Psi_\kappa$ simply by $\Psi$.
We define
$W_N$ by $W$ with $\Psi$ replaced by
$e^{ik\cdot x}$,
i.e.,
\eq{19}
W_N(x,y,|t|)=
\int\frac{\vp(k)^2}{2\omega(k)}
e^{-|t|\omega}e^{-ik\cdot (x-y)}dk.
\en
Then
\eq{81}
W_N(x,y,|t|)=
\frac{1}{4\pi^2}
 \int dX\int dY
\frac{{\wchi (X)}\wchi (Y)}
{|(X-x)-(Y-y)|^2+|t|^2}.
\en
Note that,
if $\d
\int \frac{\vp(k)^2}{\omega(k)^3}dk <\infty$,
then
$$0 \leq \sup_T \int_{-T}^0 ds \int_0^T dt W_N(x,y,|s-t|) <
\half \int \frac{\vp(k)^2}{\omega(k)^3}dk
$$
by \kak{65}.
It is however not the case
when
$\d \int \frac{\vp(k)^2}{\omega(k)^3}dk=\infty$.
This proves the following:

\bt{main1}
Suppose
Assumptions \ref{80},
 \ref{sa}
 and \ref{shortw}.
Assume $\kappa \leq \kappa_0$ and
\eq{k23}
\frac{1}{\AAA +1}+\kappa C_0 (\kappa C_0 +2)<1,
\en
where $\kappa_0$ and $C_0$ are given in
Lemma  \ref{ikebe2}.
Then $H$ has no ground state.
\et
\proof
Note that, by \kak{k23}, one can take
$0 < \lambda < 1$ such that
	\[ \frac{1}{\AAA + 1} < \lambda < 1
- \kappa C_0 (\kappa C_0 +2). \]
It is enough to show
\kak{63}, namely there exists $\varrho(T)$ such that
\eq{69}
1_{A_T}
\int_{-T}^0 ds \int_0^T dt
 \int dX\int dY
\frac{ \ov{(\Psi_{X_s}\vp)^\vee}(X)
 (\Psi_{X_t} \vp)^\vee(Y)
}
{|X-Y|^2+|s-t|^2}>\varrho(T)
\en
and $\varrho(T)\rightarrow \infty$ as $T\rightarrow \infty$.
By \kak{68} it follows that
$$\sup_{x,y,k}|
\ov{\Psi(k,x)}
\Psi(k,y)-
e^{-ikx} e^{iky}|
\leq
\kappa C_0 (\kappa C_0 +2).$$
Then
$$W(X_s,X_t,|s-t|)\geq
W_N(X_s,X_t, |s-t|)-
\kappa C_0 (\kappa C_0 +2)W_0(|t-s|),$$
where
$$W_0(|T|)=
\int\frac{\vp(k)^2}{2\omega(k)}
e^{-|T|\omega(k)}dk.$$
By \cite{lms}
on $A_T$,
\eqnn
&&\int_{-T}^0 ds \int_0^T dt
W_N(X_s,X_t,|s-t|)\non \\
&&\label{69-1}
\geq
\frac{1}{4\pi^2}
\int dXdY\wchi (X)\wchi (Y)
\log\lk
\frac
{8T^{2\la}+|X+Y|^2+T^2}
{8T^{2\la}+2|X+Y|^2}
\rk.
\ennn
Note that $\wchi\geq0$.
While
$\int_{-T}^0 ds\int_0^T dt
W_0(|t-s|)$
can be
computed as
\eqn
&&
\int_{-T}^0 ds\int_0^T dt
W_0(|t-s|)\\
&&
=
\frac{1}{4\pi^2}
\int dX\int dY
\wchi (X)\wchi (Y)
\log
\lk
\frac{(|X-Y|^2+T^2)^2}
{|X-Y|^2(|X-Y|^2+4T^2)}
\rk\\
&&
+
\frac{1}{\pi^2}
\int dX\int dY
\wchi (X)\wchi (Y)
\frac{T}{|X-Y|}
\lk
\arctan
\frac{2T}{|X-Y|}-
\arctan
\frac{T}{|X-Y|}
\rk.
\enn
The second term on the right hand side
above is uniformly
bounded by some constant $K$
with respect to $T$.
Then
\eqnn
&&
\kappa C_0 (\kappa C_0 +2)
\int_{^T}^0 ds\int _0^T dt
W_0(|t-s|)\non\\
&&
\label{69-2}
\leq
\frac{1}{4\pi^2}
\int dX\int dY
\wchi (X)\wchi (Y)
\log
\lk
\frac{(|X-Y|^2+T^2)^2}
{|X-Y|^2(|X-Y|^2+4T^2)}
\rk^{\kappa C_0 (\kappa C_0 +2)}+K.\non\\
&&
\ennn
By \kak{69-1} and \kak{69-2}
we obtain
\eq{u4}
W\geq
\frac{1}{4\pi^2}
\int\!\! dX\int\!\! dY \wchi(X)\wchi(Y) \log\lk\frac{\frac
{8T^{2\la}+|X+Y|^2+T^2}
{8T^{2\la}+2|X+Y|^2}
}{
\lk
\frac{(|X-Y|^2+T^2)^2}
{|X-Y|^2(|X-Y|^2+4T^2)}
\rk^{\kappa C_0(\kappa C_0+2)}
}\rk
-\kappa C_0(\kappa C_0+2)K
.
\en
Since $\la<1$,
$$\log\lk\frac{\frac
{8T^{2\la}+|X+Y|^2+T^2}
{8T^{2\la}+2|X+Y|^2}
}{
\lk
\frac{(|X-Y|^2+T^2)^2}
{|X-Y|^2(|X-Y|^2+4T^2)}
\rk^{\kappa C_0(\kappa C_0+2)}
}\rk
\sim \log T^{2(1-\la-\kappa C_0(\kappa C_0+2))}
$$
as $T\rightarrow \infty$,
and
$\la+\kappa C_0 (\kappa C_0 +2)<1$,
the right hand side of \kak{u4} diverges.
Then the theorem follows.
\qed

\section{The number of bosons in ground state}
In this
section we suppose
Assumptions \ref{80}, \ref{ge}  and \ref{sa}, but we do {\it not} assume $\check \chi\geq0$.
Moreover
we suppose
 the following assumption holds:
\begin{assumption}
\label{mimi}
Suppose that
(1)
$\int\frac{\vp(k)^2}{\omega(k)^3}dk <\infty$ and
(2)
 $H$ has a ground state $\gr$ such that
$\gr>0$.
\end{assumption}
Under Assumption \ref{mimi}
it follows that $\grt\rightarrow \gr$ strongly
as $T\rightarrow \infty$.
We have the proposition below.
\bp{hiroshima1}
It follows that
\eq{29}
(\grt, e^{-\beta N}\grt)=
\ET{e^{
-\g^2(1-e^{-\beta})
\WT}}.
\en
\ep
\proof
By Theorem \ref{fir2} we have
$$(\grt, e^{-\beta N}\grt)=
\frac{1}{Z_T}
\ix{e^{(\g^2/2)
\left\|
\int_{-T}^0
\phi_{r,0}(\tvp(X_r))dr
+
\int_0^T\phi_{r,\beta}(\tvp(X_r))dr
\right\|^2}}.
$$
Since
$$(\phi_{s,0}(f),\phi_{t,\beta}(g))=\half e^{-\beta}\int e^{-|t-s|\omega}\ov{\hat f(k)} \hat g(k) dk,$$
we have
\eqn
&&\left\|
\int_{-T}^0
\phi_{r,0}(\tvp(X_r))dr
+
\int_0^T\phi_{r,\beta}(\tvp(X_r))dr
\right\|^2\\
&&=
\int_{-T}^0ds
\int_{-T}^0dt W
+
\int_0^T ds
\int_0^T dt W+
e^{-\beta}
\lk
\int_{-T}^0 ds
\int _0^T   dt W +
\int_0^T    ds
\int_{-T}^0 dtW \rk\\
&&=
\int_{-T}^T ds
\int_{-T}^T dt W
+2(e^{-\beta}-1)
\int_{-T}^0 ds\int _0^T dt W.
\enn
Then the proposition follows.
\qed

Note that
\eq{30-1}
\WT\leq \half
\int\frac{\vp(k)^2}{\omega(k)^3}dk<\infty.
\en
Let
$g(\beta)=(\grt, e^{-\beta N}\grt)$.
Thus we have a lemma below:
\bl{lemma1}
For each $0<T$.
(1)
$g$ can
be analytically continued to the hole
complex plane $\CC$;
(2) $\grt\in D(e^{+\beta N})$
for all $\beta\in\CC$;
(3) \kak{29} holds true for all
$\beta\in\CC$.
\el
\proof
The proof is parallel with \cite{h03}.
Let $\Pi_+=\{z\in\CC|\Re z>0\}$ and $\Pi_-=\CC\setminus \Pi_+$.
Set
$$g(\beta)=\ET{e^{-\g^2(1-e^{-\beta})\WT}}.$$
It is easily seen that
$g(\beta)$ can be  analytically continued into the hole complex plane
$\C $ in $\beta$.
We denote its analytic continuation by $\tilde g$.
Let $\beta_0\in \Pi_+$   be such that
$\Re \beta_0=\epsilon$
with some  $\epsilon>0$.
Fix an arbitrary $R$ such that $R>\epsilon$.  We see that
\begin{equation}
\label{abb}
\tilde g(\beta)=\sum_{n=0}^\infty (\beta-\beta_0)^nb_n(\beta_0)
\end{equation}
for
$\beta\in U:=\{z\in\C \;|\; |\beta_0-z|<R\}$,
and
(\ref{abb}) absolutely converges.
Let $\nu(d\rho )$ denote the spectral projection of $N$
with respect to $\grt$.
Note that $g(\beta)$ is analytic in the interior of $\Pi_+$.
Then
\eq{kan}
g(\beta)=\int_0^\infty e^{-\beta\rho}\nu(d\rho)=\sum_{n=0}^\infty(\beta-\beta_0)^n\frac{1}{n!}\int_0^\infty (-\rho )^n e^{-\beta_0 \rho  }\nu(d\rho )
\en
for
$\beta$ so that $|\beta-\beta_0|<\epsilon$.
Since $g(\beta)=\tilde g(\beta)$ for $\beta$ such that $|\beta-\beta_0|<\epsilon$, we see together with \kak{kan} that
\eq{kan2}
b_n(\beta_0)=\frac{1}{n!}\int_0^\infty (-\rho )^n e^{-\beta_0 \rho  }\nu(d\rho ).
\en
Substituting \kak{kan2} into the expansion of $\tilde g$ in \kak{abb},
we have
\begin{equation}
\label{g}
\tilde g(\beta)=\sum_{n=0}^\infty(\beta_0-\beta)^n
\frac{1}{n!}\int_0^\infty (-\rho) ^n e^{-\beta_0\rho  }\nu(d\rho )
\end{equation}
for $\beta\in U$.
In particular the right-hand side of
 (\ref{g}) absolutely converges for $\beta\in U$,
 and
$U\cap \Pi_-\not=\emptyset$ by $R>\epsilon$,
and,
  for $\beta\in\RR\cap U\cap \Pi_-$,
by Fatou's lemma we have for any $M>0$,
\begin{eqnarray*}
\int_0^M e^{-\beta  \rho  }\nu(d\rho )
\leq
\sum_{n=0}^\infty|\beta_0-\beta|^n\frac{1}{n!}\int_0^\infty \rho ^n
e^{-\beta_0\rho  }\nu(d\rho )<\infty.
\end{eqnarray*}
Thus
$
\int_0^\infty e^{-\beta  \rho  }\nu(d\rho )<\infty$ follows
for $\beta\in \RR\cap U\cap \Pi_-$.
This
 implies that
$\gr\in D(e^{-(\beta/2) N })$ and
\kak{29} holds for
$\beta\in \RR\cap U\cap \Pi_-$.
Since $R$ is an arbitrary large number,  we get \kak{29}
for all $\beta\in\C $.
\qed
By this proposition
 the moment $(\gr, N^m\gr)$
can be derived by
\eq{30}
(\grt, N^m\grt)=(-1)^m \frac{d^m}{d\beta^m}
(\grt, e^{-\beta N}\grt)
\lceil_{\beta=0}.
\en
\bl{hiroshima3}
{\bf (Pull through formula)}
It follows that
\eq{31}
(\gr, N\gr)=
\frac{\g^2}{2}
\int dk \frac{\vp(k)^2}{\omega(k)}
\lk
\Psi(k,\cdot)\gr,
(\ov H+\omega(k))^{-2}
\Psi(k,\cdot) \gr\rk ,
\en
where $\ov H= H-\is (H)$.
\el
\proof
From
\eq{32}
(\grt, N\grt)=\ET{{\g^2}\WT}
\en
it follows that
$$
(\grt, N\grt)=
\frac{\g^2}{2}
\int dk
\frac{\vp(k)^2}{\omega(k)}
\int_{-T}^0 ds\int_0^T dt e^{-|t-s|\omega}
\ET{
\ov{\Psi(k,X_s)}\Psi(k,X_t)}.$$
Generally
it can be obtained
that for bounded $f$ and $g$,
\eq{90}
\ET{f(X_s)g(X_t)}=
(e^{-sH}\grt, fe^{-(t-s)H}g e^{+tH}\grt),\quad t\geq s.
\en
This can be proven directly by the Trotter product formula.
Then since
\eq{72}
\ET{
\ov{\Psi(k,X_s)}\Psi(k,X_t)}=
\lk
\Psi(k,\cdot)e^{-sH}\grt,
e^{-(t-s) H}
\Psi(k,\cdot) e^{+tH}\grt\rk,
\en
we have
\eqn
&&(\grt, N\grt)\\
&&=
\frac{\g^2}{2}
\int dk \frac{\vp(k)^2}{\omega(k)}
\int_{-T}^0ds\int_0^Tdt e^{-|t-s|\omega}
\lk
\Psi(k,\cdot)e^{-sH}\grt,
e^{-(t-s) H}
\Psi(k,\cdot) e^{+tH}\grt\rk.
\enn
Since \kak{32} yields that
$$\|N^\han\grt\|
\leq
\frac{\g^2}{2}
\int\frac{\vp(k)^2}{\omega(k)^3}dk
<\infty,$$
there exists a subsequence
$T'$ such that
\eq{74}
s-\lim_{T'\rightarrow\infty}
N^\han \gr^{T'}=N^\han \gr.
\en
Let us reset $T$ for $T'$.
By \kak{72}
$$|
\lk
\Psi(k,\cdot)
e^{-sH}\grt,
e^{-(t-s) H}
\Psi(k,\cdot)
e^{+tH}\grt\rk|\leq
\sup_{k,x}
|\Psi(k,x)|^2<\infty$$
and
$$\lim_{T\rightarrow\infty}
\lk
\Psi(k,\cdot)e^{-sH}\grt,
e^{-(t-s) H}
\Psi(k,\cdot) e^{+tH}\grt\rk
=
\lk
\Psi(k,\cdot)\gr,
e^{-(t-s) \ov{H}}
\Psi(k,\cdot) \gr\rk.$$
By the dominated convergence theorem
we have
\eqnn
&&
\lim_{N\rightarrow \infty}
\int dk \frac{\vp(k)^2}{2\omega(k)}
\int_{-T}^0ds
\int_0^T dt
e^{-|t-s|\omega}
\lk
\Psi(k,\cdot)e^{-sH}\grt,
e^{-(t-s) H}
\Psi(k,\cdot) e^{+tH}\grt\rk\non \\
&&
\hspace{1cm}
\label{75}
=
\int dk\frac{\vp(k)^2}{2\omega(k)}
\int_{-\infty}^0
ds
\int_0^\infty
dt
e^{-|t-s|\omega}
\lk
\Psi(k,\cdot)\gr,
e^{-(t-s) \ov{H}}
\Psi(k,\cdot) \gr
\rk.
\ennn
The right hand side above is identical with
$$
\int dk \frac{\vp(k)^2}{2\omega(k)}
\lk
\Psi(k,\cdot)\gr,
(\ov H+\omega(k))^{-2}
\Psi(k,\cdot) \gr\rk.$$
By \kak{74} and \kak {75} the lemma follows.
\qed

\bt{main2}
Set $R=\int\frac{\vp(k)^2}{\omega(k)^3}dk
$.
Suppose that $(\Psi(0,\cdot)\gr,\gr)\not=0$.
Then
\eq{k30-1}
\lim_{R\rightarrow\infty}
(\gr, N\gr)=\infty.
\en
\et

\begin{example}{\rm
Assume that
$\vm =\kappa w$ satisfies
Assumption \ref{shortw}.
Then
$|1-\Psi(0,x)| \leq \kappa C_0$ holds
by Lemma \ref{ikebe2}.
It yields that
	\[ |(\Psi(0,\cdot)\gr,\gr) -1| \leq \kappa C_0. \]
Thus $(\Psi(0,\cdot)\gr,\gr)\not=0$ holds
for sufficiently small $\kappa$.
}\end{example}

{\it Proof of Theorem \ref{main2}}

By Lemma \ref{hiroshima3} we have
\eq{83}
(\gr, N\gr)=
\frac{\g^2}{2}
\int dk
\frac{\vp(k)^2}{\omega(k)^3}
\lk
\Psi(k,\cdot)\gr,
\omega(k)^2
(\ov H+\omega(k))^{-2}
\Psi(k,\cdot) \gr\rk.
\en
We can see that
\begin{eqnarray*}
&&\lim_{|k|\rightarrow 0}
\left|
(\Psi(k,\cdot)\gr, \omega(k)^{2}
(
\ov H+\omega(k))^{-2}\Psi(k,\cdot)\gr)\right.\\
&&
\hspace{1cm}
-
\left.
(\Psi(0,\cdot)
\gr, \omega(k)^{2}
(
\ov H+\omega(k))^{-2}\Psi(0,\cdot)\gr)\right|=0.
\end{eqnarray*}
Let $P_{\rm g}$ (resp. $P_{\rm g}^\perp$) denote the projection to the ground state $\ker \bar H$
(resp. the orthogonal complement $(\ker \bar H)^\perp$ of $\ker \bar H$).
 We have
\begin{eqnarray*}
&&\hspace{-3cm}
(\Psi(0,\cdot)\gr,
\omega(k)^{2}(\ov H+\omega(k))^{-2}\Psi(0,\cdot)\gr)\\
&=&
(\Psi(0,\cdot)\gr,
\omega(k)^{2}(\ov H+\omega(k))^{-2}
(P_{\rm g}+P_{\rm g}^\perp)
\Psi(0,\cdot)\gr)
\end{eqnarray*}
Then
$$\lim_{|k|\rightarrow 0}
(\Psi(0,\cdot)\gr,
\omega(k)^{2}(\ov H+\omega(k))^{-2} P_{\rm g}
\Psi(0,\cdot)\gr)=
|(\gr, \Psi(0,\cdot)\gr)|^2$$
and
$$\lim_{|k|\rightarrow 0}
(\Psi(0,\cdot)\gr,
\omega(k)^{2}(\ov H+\omega(k))^{-2} P_{\rm g}^\perp
\Psi(0,\cdot)\gr)=0.$$
Then we conclude that
\eq{33-1}
\lim_{|k|\rightarrow 0}
(\Psi(k,\cdot)\gr,
\omega(k)^2 (\ov H+\omega(k))^{-2}
\Psi(k,\cdot)\gr)=
|(\Psi(0,\cdot)\gr,\gr)|^2.
\en
Set
$A=|(\Psi(0,\cdot)\gr,\gr)|^2>0$.
Then
$$
A-\delta<
(\Psi(k,\cdot)\gr,
\omega(k)^2 (\ov H+\omega(k))^{-2}
\Psi(k,\cdot)\gr)$$
for $|k|<\epsilon$ with some sufficiently small $\epsilon>0$.
Then
we have the bound
\eq{34}
(A-\delta) \frac{\g^2}{2}\int_{|k|<\epsilon}
\frac{\vp(k)^2}{\omega(k)^3}dk
+\frac{\g^2}{2}\int_{|k|\geq \epsilon} \frac{\vp(k)^2}{\omega(k)^3} dk
\leq (\gr, N\gr)
\en
with some positive $b$.
 Thus as $R\rightarrow \infty$,
$(\gr, N\gr)$ goes to infinity.
Then the proof is complete.
\qed

\noindent {\bf Acknowledgments:}
FH acknowledges support of Grant-in-Aid for
Science Research (B) 20340032 from JSPS
and is thankful to the hospitality
of IHES, Bures-sur-Yvette, and
Universit\'e de Paris XI,
where part of this work has been done.
AS acknowledges the financial support of Global COE program {\it Mathematics-for-Industry} in Kyushu university.

\end{document}